\documentclass[twoside,twocolumn,9pt]{article}
\usepackage{extsizes}
\usepackage[super,sort&compress,comma]{natbib} 
\usepackage[version=3]{mhchem}
\usepackage[left=1.5cm, right=1.5cm, top=1.785cm, bottom=2.0cm]{geometry}
\usepackage{balance}
\usepackage{mathptmx}
\usepackage{sectsty}
\usepackage{graphicx} 
\usepackage{lastpage}
\usepackage[format=plain,justification=justified,singlelinecheck=false,font={stretch=1.125,small,sf},labelfont=bf,labelsep=space]{caption}
\usepackage{float}
\usepackage{cancel}
\usepackage[normalem]{ulem}
\usepackage{fancyhdr}
\usepackage{fnpos}
\usepackage[english]{babel}
\addto{\captionsenglish}{%
  \renewcommand{\refname}{Notes and references}
}
\usepackage{array}
\usepackage{droidsans}
\usepackage{charter}
\usepackage[T1]{fontenc}
\usepackage[usenames,dvipsnames]{xcolor}
\usepackage{setspace}
\usepackage[compact]{titlesec}
\usepackage{hyperref}
\hypersetup{
    colorlinks=true,
    linkcolor=black,
    citecolor=blue,
    urlcolor=blue
}

\usepackage{epstopdf}

\definecolor{cream}{RGB}{222,217,201}

\begin{document}

\pagestyle{fancy}
\thispagestyle{plain}
\fancypagestyle{plain}{
\renewcommand{\headrulewidth}{0pt}
}

\makeFNbottom
\makeatletter
\renewcommand\LARGE{\@setfontsize\LARGE{15pt}{17}}
\renewcommand\Large{\@setfontsize\Large{12pt}{14}}
\renewcommand\large{\@setfontsize\large{10pt}{12}}
\renewcommand\footnotesize{\@setfontsize\footnotesize{7pt}{10}}
\makeatother

\renewcommand{\thefootnote}{\fnsymbol{footnote}}
\renewcommand\footnoterule{\vspace*{1pt}%
\color{cream}\hrule width 3.5in height 0.4pt \color{black}\vspace*{5pt}} 
\setcounter{secnumdepth}{5}

\makeatletter 
\renewcommand\@biblabel[1]{#1}            
\renewcommand\@makefntext[1]%
{\noindent\makebox[0pt][r]{\@thefnmark\,}#1}
\makeatother 
\renewcommand{\figurename}{\small{Fig.}~}
\sectionfont{\sffamily\Large}
\subsectionfont{\normalsize}
\subsubsectionfont{\bf}
\setstretch{1.125} 
\setlength{\skip\footins}{0.8cm}
\setlength{\footnotesep}{0.25cm}
\setlength{\jot}{10pt}
\titlespacing*{\section}{0pt}{4pt}{4pt}
\titlespacing*{\subsection}{0pt}{15pt}{1pt}



\fancyfoot{}
\fancyfoot[LO,RE]{\vspace{-7.1pt}\includegraphics[height=9pt]{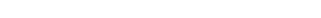}}
\fancyfoot[CO]{\vspace{-7.1pt}\hspace{13.2cm}\includegraphics{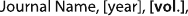}}
\fancyfoot[CE]{\vspace{-7.2pt}\hspace{-14.2cm}\includegraphics{head_foot/RF}}
\fancyfoot[RO]{\footnotesize{\sffamily{1--\pageref{LastPage} ~\textbar  \hspace{2pt}\thepage}}}
\fancyfoot[LE]{\footnotesize{\sffamily{\thepage~\textbar\hspace{3.45cm} 1--\pageref{LastPage}}}}
\fancyhead{}
\renewcommand{\headrulewidth}{0pt} 
\renewcommand{\footrulewidth}{0pt}
\setlength{\arrayrulewidth}{1pt}
\setlength{\columnsep}{6.5mm}
\setlength\bibsep{1pt}

\makeatletter 
\newlength{\figrulesep} 
\setlength{\figrulesep}{0.5\textfloatsep} 

\newcommand{\topfigrule}{\vspace*{-1pt}%
\noindent{\color{cream}\rule[-\figrulesep]{\columnwidth}{1.5pt}} }

\newcommand{\botfigrule}{\vspace*{-2pt}%
\noindent{\color{cream}\rule[\figrulesep]{\columnwidth}{1.5pt}} }

\newcommand{\dblfigrule}{\vspace*{-1pt}%
\noindent{\color{cream}\rule[-\figrulesep]{\textwidth}{1.5pt}} }
\newcommand{\new}[1]{\textcolor{cyan} { #1}}

\makeatother
\twocolumn[
  \begin{@twocolumnfalse}
{\includegraphics[height=30pt]{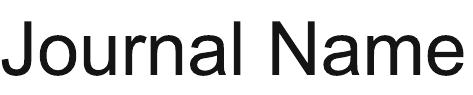}\hfill\raisebox{0pt}[0pt][0pt]{\includegraphics[height=55pt]{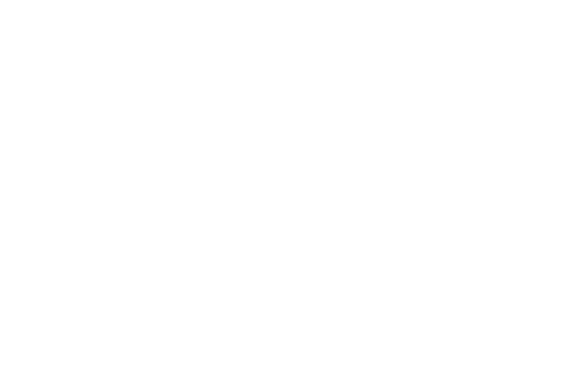}}\\[1ex]
\includegraphics[width=18.5cm]{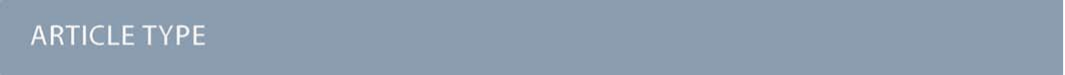}}\par
\vspace{1em}
\sffamily
\begin{tabular}{m{4.5cm} p{13.5cm} }


\includegraphics{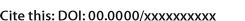} & \noindent\LARGE{\textbf{AC Field-driven orientational crossover and energy dissipation in suspended magnetic nanoparticles.
}}
\\
\vspace{0.3cm} & \vspace{0.3cm} \\


& \noindent\large{Iago L\'opez-V\'azquez, $^{\ast}$\textit{$^{a,b}$}, Siraj Ul Haq\textit{$^{a,b}$}, Kazuya Okada\textit{$^{c}$}, Sergiu Ruta\textit{$^{d}$}, Roy W. Chantrell\textit{$^{e}$}, {\`O}scar Iglesias\textit{$^{f}$} and David Serantes\textit{$^{a,b}$} } \\



\includegraphics{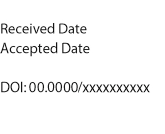} & \noindent\normalsize{
By combining the Landau--Lifshitz--Gilbert equation with Brownian rotational dynamics of magnetic nanoparticles (MNPs), we theoretically investigate the role of particle rotation through easy-axis reorientation in magnetic fluid hyperthermia (MFH). 
Our results reveal a field-driven crossover in the stationary orientation of the easy axes, from predominantly perpendicular to predominantly parallel or antiparallel to the applied field as the field amplitude increases. 
Although the precise crossover field depends on particle size and excitation frequency, it occurs at approximately $0.5H_k$, where $H_k$ is the uniaxial anisotropy field. 
These orientational regimes are directly linked to the underlying microscopic dynamics and the associated MFH performance through the occurrence of switching and non-switching hysteresis cycles, predominantly associated with N\'eel magnetization reversal and Brownian particle rotation, respectively.
The relative importance of these dissipation mechanisms also depends on frequency: at $f=1$ MHz, Brownian heating dominates at low field amplitudes, whereas N\'eel heating dominates at high fields. By contrast, at $f=100$ kHz, both contributions remain comparable over most of the investigated field range.}

\end{tabular}

\end{@twocolumnfalse} \vspace{0.6cm}

]



\renewcommand*\rmdefault{bch}\normalfont\upshape
\rmfamily
\section*{}
\vspace{-1cm}


\footnotetext{%
\textit{
$^{a}$~
Departamento de F\'isica Aplicada,
Universidade de Santiago de Compostela,
15782 Santiago de Compostela,
Galicia, Spain. Email: iago.lopez@usc.es
} }

\footnotetext{%
\textit{
$^{b}$~
Instituto de Materiais (iMATUS),
Universidade de Santiago de Compostela,
15782 Santiago de Compostela,
Galicia, Spain.
} }

\footnotetext{%
\textit{
$^{c}$~
Department of Mechanical Engineering,
Saitama Institute of Technology,
Fukaya, Japan.
} }

\footnotetext{\textit{$^{d}$~College of Business, Technology and Engineering, Sheffield Hallam University, UK.}}

\footnotetext{%
\textit{
$^{e}$~
School of Physics, Engineering and Technology,
University of York,
York, UK.
} }

\footnotetext{%
\textit{
$^{f}$~
Departamento de F\'isica de la Mat\`eria Condensada,
Universitat de Barcelona and IN2UB,
c/ Mart\'{\i} i Franqu\`es 1,
08028 Barcelona,
Catalunya, Spain.
} }

\section{Introduction}

Magnetic fluid hyperthermia (MFH) relies on the capability of magnetic nanoparticles (MNPs) to dissipate energy when driven by an alternating magnetic field \cite{perigo2015fundamentals,gavilan2025magnetic}, commonly quantified through the \textit{Specific Loss Power} (SLP) or \textit{Specific Absorption Rate} (SAR) \cite{wells2021challenges}. Many theoretical descriptions focus exclusively on the internal dynamics of the magnetic moment \cite{behbahani2022micromagnetic,ruta2015unified,failde2024understanding}, thereby assuming that the MNPs remain physically blocked and that their response is entirely governed by N\'eel relaxation; several experiments support this approach \cite{beola2020,soukup2015situ,di2014magnetic}. 
However, a growing body of experimental evidence shows that heating efficiency depends strongly on the viscosity of the surrounding medium, indicating that mechanical degrees of freedom can play an essential role under realistic conditions \cite{saville2014formation,phong2017study,cabrera2017unraveling}. This viscosity dependence becomes particularly relevant in biological environments, where effective viscosities may differ from that of water by several orders of magnitude and  vary across length scales \cite{kwapiszewska2020nanoscale}.

To understand this scenario, an essential component to address is the coupling between magnetization dynamics and particle motion, since the relevant translational and rotational relaxation times depend explicitly on the viscosity of the surrounding medium. 
Although these processes are closely coupled, their influence on MFH
performance can, as a first approximation, be analyzed separately at the single-particle and collective-particle levels, respectively. 

The first level corresponds to the time-dependent rotation of the particle body, which we describe in terms of the rotation of the anisotropy easy axes. Besides producing energy dissipation through rotational Brownian losses \cite{sanchez2009rotational,suzuki2021behaviour}, 
particle rotation also modifies the orientation of the easy axis, thereby changing the anisotropy landscape that governs magnetization reversal and the associated heat release \cite{serantes2018anisotropic,simeonidis2016situ}.
The second level corresponds to changes in the local energy barriers 
arising from the evolution of dipolar interparticle interactions during the slow structural reorganization of the colloid, for example through chain formation, which can also modify heat production \cite{serantes2014multiplying}.

In the present work, we focus on the first of these mechanisms and investigate the role of easy-axis reorientation in MFH performance. Understanding the single-particle response under AC magnetic fields is also of interest beyond MFH. In particular, it may be relevant in other contexts such as magnetogenetics \cite{del2022magnetogenetics,latypova2024magnetogenetics}, a field in which the physical mechanisms underlying the reported biological responses remain under debate \cite{meister2016physical} and where reproducibility issues have also been highlighted \cite{kole2020assessing,wang2020revaluation}. In this broader context, an accurate description of particle-level behaviour, including both heating and mechanically induced rotation, may help clarify whether MNPs can generate not only thermal effects but also localized mechanical stresses at the nanoscale. Another application area in which this description may be relevant is \textit{magnetic particle imaging} (MPI) \cite{healy2022clinical,velazquez2025advances}, where changes in magnetic relaxation directly influence signal generation and image quality \cite{utkur2017relaxation}.

The combined role of magnetization dynamics and particle rotation in magnetic heating was, to the best of our knowledge, first addressed by Mamiya and Jeyadevan in \cite{mamiya2011hyperthermic}, where they investigated the time evolution of the easy-axis orientation under AC field excitation and examined how rotational degrees of freedom influence the magnetic response and the associated heating. This line of research was subsequently extended by other authors, including Usov and Liubimov \cite{usov2012dynamics}, Reeves and Weaver \cite{reeves2014nonlinear}, Ota and Takemura \cite{ota2017evaluation}, and, more recently, Wolfschwenger \textit{et al.}  \cite{wolfschwenger2024molecular} and Durhuus \textit{et al.}  \cite{durhuus2024conservation}. 

Motivated by these previous studies, 
here we investigate the coupled evolution of the magnetic moment and the particle orientation in a viscous environment within a Landau--Lifshitz--Gilbert--Brownian Dynamics (LLG--BD) framework \cite{okada2025proposal}. 
Our objective is not only to quantify the influence of particle rotation on magnetic hyperthermia, but also to establish a direct connection between easy-axis reorientation, magnetization-reversal dynamics, and energy dissipation. In particular, we identify distinct orientational regimes under AC excitation and relate them to the occurrence of switching and non-switching hysteresis cycles and their corresponding contributions to heat generation.
As a first approximation, we restrict the analysis to non-interacting particles with an effective uniaxial anisotropy, so that the magnetic and rotational degrees of freedom remain coupled through the anisotropy energy.

The manuscript is organized as follows. In Section~\ref{sec:model}, we introduce the computational model, the simulation framework, and the main parameters used throughout the study. 
Section~\ref{sec:easy_axis_reorientation} investigates the reorientation of the nanoparticle easy axes under AC field excitation, including the transient evolution of the orientational distribution (Section~\ref{sec:transient}), the stationary orientational states (Section~\ref{sec:stationary_alignment}), and their dependence on particle size, viscosity, and excitation frequency (Section~\ref{sec:size_viscosity_frequency}).
Establishing these reorientation regimes is an essential step toward understanding the long-time heating behaviour, since the orientation of the anisotropy easy axis determines the effective energy landscape experienced by the magnetization and therefore influences its dynamical response to the applied field. 

The connection between reorientation, switching dynamics, and energy dissipation is addressed in Section~\ref{sec:switching}, where we analyze hysteresis cycles and classify the trajectories into switching and non-switching regimes. The former exhibit abrupt magnetization reversals and are therefore associated with N\'eel-like behaviour, whereas in the latter the magnetization evolves smoothly while remaining approximately aligned with the instantaneous easy axis as the particle rotates, which we refer to as Brownian-like behaviour. Finally, we relate these dynamical regimes to their corresponding hysteresis losses. 

\section{Model and simulation framework}
\label{sec:model}

To isolate the coupled magnetic and rotational response of
individual particles, we consider a monodisperse ensemble of
non-interacting MNPs dispersed in a fluid with
dynamic viscosity $\eta$. The particles are subjected to an alternating
magnetic field of amplitude $H_{\max}$ and frequency $f$,

\begin{equation}
\mathbf{H}_{\mathrm{app}}(t)
=
H_{\max}\sin(2\pi f t)\,\hat{\mathbf{H}},
\label{eq:Happ}
\end{equation}
where $\hat{\mathbf{H}}$ is a unit vector along the field direction.

Each nanoparticle is described within the macrospin approximation and
is characterized by a magnetic moment of constant magnitude
$m=M_sV$, where $M_s$ is the saturation magnetization and $V$ is the
magnetic volume. For the spherical magnetic cores considered here,
$V=\pi d^3/6$, where $d$ is the particle diameter.

\begin{figure}[!t]
    \centering
    \includegraphics[width=0.6\columnwidth]{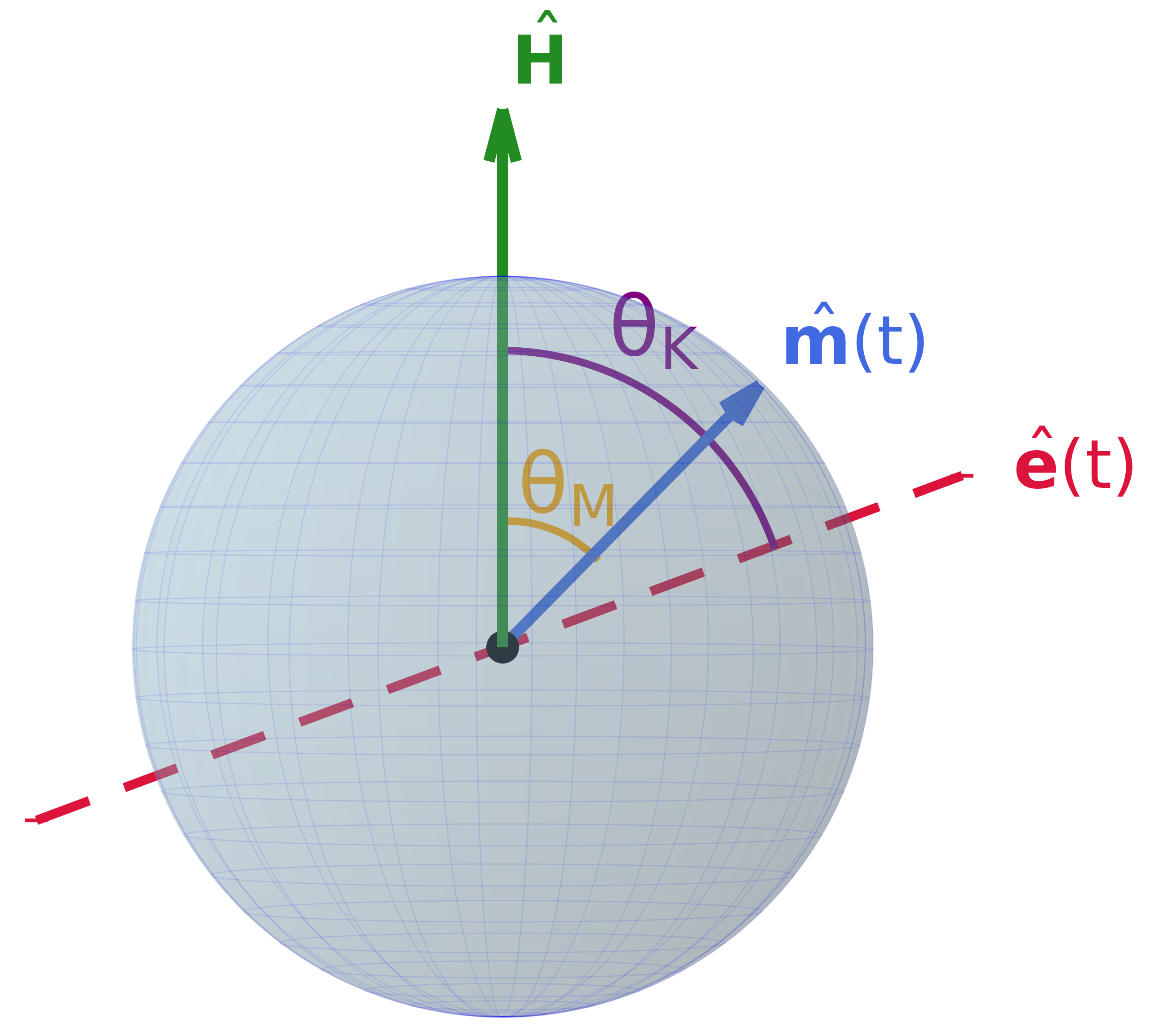}
    \caption{Schematic representation of the angles considered in the
    analysis. $\theta_M$ and $\theta_K$ are, respectively, the angles
    formed by the magnetic moment $\hat{\mathbf{m}}(t)$ and the easy
    axis $\hat{\mathbf{e}}(t)$ with the applied-field direction
    $\hat{\mathbf{H}}$.}
    \label{fig:scheme}
\end{figure}

The state of each particle is therefore described by two time-dependent unit vectors: the magnetization direction $\hat{\mathbf{m}}(t)$ and the body-fixed uniaxial easy-axis direction $\hat{\mathbf{e}}(t)$, as illustrated in Figure~\ref{fig:scheme}.

\subsection{Coupled LLG--Brownian dynamics}
\label{subsec:coupled_dynamics}

The magnetic and rotational degrees of freedom are coupled through the
uniaxial anisotropy energy,

\begin{equation}
U_{\mathrm{ani}}
=
-K_{\mathrm{u}}V
\left(
\hat{\mathbf{m}}\cdot\hat{\mathbf{e}}
\right)^2,
\label{eq:Uani_main}
\end{equation}
where $K_{\mathrm{u}}$ is the effective uniaxial anisotropy constant.

The stochastic evolution of the magnetic moment is described by the
Landau--Lifshitz--Gilbert (LLG) equation
\cite{satohintroduction,satoh2017book},
\begin{equation}
\frac{d\hat{\mathbf{m}}}{dt}
=
-\frac{\gamma}{1+\alpha^2}
\left[
\hat{\mathbf{m}}\times\mathbf{H}_{\mathrm{eff}}
+
\alpha\,
\hat{\mathbf{m}}\times
\left(
\hat{\mathbf{m}}\times\mathbf{H}_{\mathrm{eff}}
\right)
\right],
\label{eq:LLG_main}
\end{equation}
where $\gamma$ is the gyromagnetic ratio and $\alpha$ is the Gilbert
damping parameter. The effective field is written as

\begin{equation}
\mathbf{H}_{\mathrm{eff}}
=
\mathbf{H}_{\mathrm{app}}(t)
+
\mathbf{H}_{\mathrm{ani}}
+
\mathbf{H}_{\mathrm{th}}(t),
\label{eq:Heff}
\end{equation}
where $\mathbf{H}_{\mathrm{ani}}$ is the uniaxial anisotropy field and
$\mathbf{H}_{\mathrm{th}}$ represents thermal magnetic fluctuations.

Using Eq.~\eqref{eq:Uani_main}, the anisotropy field is obtained as
\begin{equation}
\mathbf{H}_{\mathrm{ani}}
=
-\frac{1}{\mu_0M_sV}
\frac{\partial U_{\mathrm{ani}}}
{\partial\hat{\mathbf{m}}}
=
\frac{2K_{\mathrm{u}}}{\mu_0M_s}
\left(
\hat{\mathbf{m}}\cdot\hat{\mathbf{e}}
\right)
\hat{\mathbf{e}},
\label{eq:Hani_uni}
\end{equation}
where $\mu_0$ is the vacuum permeability. The corresponding uniaxial
anisotropy field scale is

\begin{equation}
H_k=\frac{2K_{\mathrm{u}}}{\mu_0M_s}.
\label{eq:Hk}
\end{equation}
Thermal magnetic fluctuations are introduced through a Gaussian
white-noise field satisfying

\begin{equation}
\begin{aligned}
\left\langle
\mathbf{H}_{\mathrm{th}}(t)
\right\rangle
&=
\mathbf{0},
\\
\left\langle
H_{\mathrm{th},i}(t)
H_{\mathrm{th},j}(t')
\right\rangle
&=
\frac{2\alpha k_BT}
{\gamma\mu_0M_sV}
\delta_{ij}\delta(t-t'),
\end{aligned}
\label{eq:Hth_statistics}
\end{equation}
where $k_B$ is Boltzmann's constant and $T$ is the temperature.

The magnetization dynamics is coupled to the physical rotation of the
particle in the surrounding viscous fluid. The rotational dynamics of
a rigid particle can be expressed through the Langevin equation
\cite{satohintroduction,satoh2017book,kim2013microhydrodynamics},

\begin{equation}
I\frac{d\boldsymbol{\omega}}{dt}
=
\boldsymbol{\Gamma}_{\mathrm{det}}
-
\xi_r\boldsymbol{\omega}
+
\boldsymbol{\Gamma}_{\mathrm{th}},
\label{eq:Langevin_rot}
\end{equation}
where $I$ is the particle moment of inertia,
$\boldsymbol{\omega}$ is its angular velocity, $\xi_r$ is the
rotational friction coefficient, and
$\boldsymbol{\Gamma}_{\mathrm{det}}$ and
$\boldsymbol{\Gamma}_{\mathrm{th}}$ are the deterministic and thermal
mechanical torques, respectively.

For nanoparticles dispersed in liquid media, inertial effects are
negligible on the timescales considered here. The rotational dynamics
can therefore be treated in the overdamped limit, for which

\begin{equation}
\xi_r\boldsymbol{\omega}
=
\boldsymbol{\Gamma}_{\mathrm{det}}
+
\boldsymbol{\Gamma}_{\mathrm{th}},
\qquad
\boldsymbol{\omega}
=
\frac{
\boldsymbol{\Gamma}_{\mathrm{det}}
+
\boldsymbol{\Gamma}_{\mathrm{th}}
}{
\xi_r
}.
\label{eq:omega_overdamped}
\end{equation}
The body-fixed easy axis evolves according to the rigid-body
kinematic relation

\begin{equation}
\frac{d\hat{\mathbf{e}}}{dt}
=
\boldsymbol{\omega}\times\hat{\mathbf{e}}.
\label{eq:BD_main}
\end{equation}
A rigid-body rotation changes $\hat{\mathbf{e}}$ and generates the
deterministic anisotropy torque

\begin{equation}
\boldsymbol{\Gamma}_{\mathrm{det}}
=
\frac{\partial U_{\mathrm{ani}}}
{\partial\hat{\mathbf{e}}}
\times\hat{\mathbf{e}}
=
2K_{\mathrm{u}}V
\left(
\hat{\mathbf{m}}\cdot\hat{\mathbf{e}}
\right)
\left(
\hat{\mathbf{e}}\times\hat{\mathbf{m}}
\right).
\label{eq:Gamma_det}
\end{equation}
This torque tends to rotate the particle so that its easy axis becomes
parallel or antiparallel to the instantaneous magnetization direction.

The thermal torque is also modeled as Gaussian white noise, with
\begin{equation}
\begin{aligned}
\left\langle
\boldsymbol{\Gamma}_{\mathrm{th}}(t)
\right\rangle
&=
\mathbf{0},
\\
\left\langle
\Gamma_{\mathrm{th},i}(t)
\Gamma_{\mathrm{th},j}(t')
\right\rangle
&=
2k_BT\xi_r
\delta_{ij}\delta(t-t').
\end{aligned}
\label{eq:Gammath_statistics}
\end{equation}
We assume isotropic rotational mobility corresponding to spherical
hydrodynamic drag. For a particle with hydrodynamic radius $R_h$, the
rotational friction coefficient and rotational diffusion coefficient
are
\begin{equation}
\xi_r=8\pi\eta R_h^3,
\qquad
D_r=\frac{k_BT}{\xi_r}.
\label{eq:rotational_coefficients}
\end{equation}
When a non-magnetic coating of thickness $t_{\mathrm{nm}}$ is present,
the hydrodynamic radius can be expressed as
$R_h=R+t_{\mathrm{nm}}$, where $R$ is the magnetic core radius. No
additional coating thickness is included in the present simulations,
and therefore $R_h=d/2$.

Equations~\eqref{eq:LLG_main} and \eqref{eq:BD_main} define the coupled
LLG--Brownian dynamics model. The coupling is twofold: the anisotropy
field acting on the magnetic moment depends on the instantaneous
easy-axis direction through Eq.~\eqref{eq:Hani_uni}, while the
mechanical torque acting on the particle depends on the instantaneous
magnetization direction through Eq.~\eqref{eq:Gamma_det}. Consequently,
$\hat{\mathbf{m}}$ and $\hat{\mathbf{e}}$ evolve self-consistently
under the applied field, anisotropy, viscous damping, and magnetic and
mechanical thermal fluctuations.

Translational Brownian motion is not included because the particles are
non-interacting and the present analysis concerns only their internal
magnetization and rotational degrees of freedom.

\subsection{Numerical integration scheme}
\label{subsec:numerical_integration}

The coupled magnetic and rotational equations are integrated using the
adaptive Dormand--Prince method (DP45)
\cite{dormand1980family}. This embedded Runge--Kutta scheme provides
fourth- and fifth-order approximations within each trial time step,
allowing the local truncation error to be estimated and the time step
to be adjusted automatically.

For an ensemble containing $N_{\mathrm{part}}$ particles, the complete
state vector is written as
\begin{equation}
\mathbf{y}(t)
\equiv
\left(
\hat{\mathbf{m}}_1,\hat{\mathbf{e}}_1,
\ldots,
\hat{\mathbf{m}}_{N_{\mathrm{part}}},
\hat{\mathbf{e}}_{N_{\mathrm{part}}}
\right),
\label{eq:state_vector}
\end{equation}
and the coupled dynamics can be expressed in compact form as
\begin{equation}
\frac{d\mathbf{y}}{dt}
=
\mathbf{G}(\mathbf{y},t).
\label{eq:compact_dynamics}
\end{equation}
For a trial step of size $h$, the DP45 method evaluates the
intermediate stages
\begin{equation}
\mathbf{k}_i
=
\mathbf{G}
\left(
\mathbf{y}
+
h\sum_{j<i}a_{ij}\mathbf{k}_j,
\,
t+c_i h
\right),
\label{eq:DP_stages}
\end{equation}
and constructs the two embedded solutions
\begin{equation}
\begin{aligned}
\mathbf{y}^{(5)}
&=
\mathbf{y}
+
h\sum_i b_i^{(5)}\mathbf{k}_i,
\\
\mathbf{y}^{(4)}
&=
\mathbf{y}
+
h\sum_i b_i^{(4)}\mathbf{k}_i,
\end{aligned}
\label{eq:DP_solutions}
\end{equation}

where $(a_{ij},c_i,b_i^{(4)},b_i^{(5)})$ are the Dormand--Prince
coefficients \cite{dormand1980family}. The fifth-order solution is used
to advance the system, while the fourth-order solution provides an
estimate of the local numerical error.

The error associated with the complete ensemble is evaluated as the
maximum vector-norm difference over all particles and both dynamical
degrees of freedom,
\begin{equation}
\varepsilon
=
\max_{p=1,\ldots,N_{\mathrm{part}}}
\left\{
\left\|
\hat{\mathbf{m}}_p^{(5)}
-
\hat{\mathbf{m}}_p^{(4)}
\right\|,
\,
\left\|
\hat{\mathbf{e}}_p^{(5)}
-
\hat{\mathbf{e}}_p^{(4)}
\right\|
\right\}.
\label{eq:DP_error}
\end{equation}
A trial step is accepted when
\begin{equation}
\varepsilon\leq\mathrm{absTol},
\label{eq:step_acceptance}
\end{equation}
where $\mathrm{absTol}$ is the prescribed absolute tolerance. After
each trial step, the time increment is updated according to
\begin{equation}
h_{\mathrm{new}}
=
h\,
\min
\left[
5,\,
\max
\left(
0.2,\,
0.9
\left(
\frac{\mathrm{absTol}}{\varepsilon}
\right)^{1/5}
\right)
\right],
\label{eq:h_update}
\end{equation}
where the exponent reflects the fifth-order accuracy of the main
solution, while the numerical prefactors prevent excessively abrupt
changes in the time step \cite{hairer1993solving}. In the present
calculations, $\mathrm{absTol}=10^{-7}$.

After every accepted step, the unit-length constraints are enforced by
explicit renormalization,
\begin{equation}
\hat{\mathbf{m}}_p
\leftarrow
\frac{\hat{\mathbf{m}}_p}
{\left|\hat{\mathbf{m}}_p\right|},
\qquad
\hat{\mathbf{e}}_p
\leftarrow
\frac{\hat{\mathbf{e}}_p}
{\left|\hat{\mathbf{e}}_p\right|},
\label{eq:renormalization}
\end{equation}
for every particle $p$.

Both stochastic contributions scale as $1/\sqrt{\Delta t}$. For each
trial step of size $\Delta t=h$, one Gaussian realization of the
thermal magnetic field and one realization of the thermal mechanical
torque are generated and kept fixed throughout all Runge--Kutta
sub-stages of that trial step.

When a step is rejected and recomputed with a different time increment,
the same Gaussian random realizations are retained and their amplitudes
are rescaled according to
\begin{equation}
\mathbf{X}_{\mathrm{th}}^{\,\mathrm{new}}
=
\mathbf{X}_{\mathrm{th}}^{\,\mathrm{old}}
\sqrt{
\frac{
\Delta t_{\mathrm{old}}
}{
\Delta t_{\mathrm{new}}
}
},
\label{eq:noise_rescale}
\end{equation}
where $\mathbf{X}_{\mathrm{th}}$ denotes either
$\mathbf{H}_{\mathrm{th}}$ or
$\boldsymbol{\Gamma}_{\mathrm{th}}$. This procedure preserves the
appropriate fluctuation--dissipation balance under adaptive time
stepping and avoids the introduction of spurious correlations
\cite{leliaert2017adaptively}.

\subsection{Physical assumptions and simulation parameters}
\label{subsec:simulation_parameters}

In the present work, we restrict ourselves to particles whose magnetic anisotropy is described solely by an effective uniaxial contribution. 
This approximation was adopted to isolate the interplay between magnetization dynamics and particle rotation within the simplest coupled LLG--BD framework. However, we note that for magnetite nanoparticles the cubic magnetocrystalline anisotropy may also play a relevant role, particularly in slightly elongated particles \cite{failde2024understanding}.
Including both uniaxial and cubic contributions within the present coupled LLG--BD framework would substantially increase the complexity of the model. In this first study, we therefore focus on the purely uniaxial case.

The main parameters of the simulations are summarized in Table~\ref{tab:simulation_parameters}. Unless otherwise stated, all calculations are performed at room temperature using an ensemble of 1000 initially randomly oriented non-interacting particles with uniaxial anisotropy. 
The magnetic parameters employed ($M_s$ and $K_u$) are representative of magnetite nanoparticles.
The particle diameter, solvent viscosity, excitation frequency, and reduced field amplitude are varied depending on the specific analysis presented in each subsection. In the following we focus first on the easy-axis alignment induced by an AC field and follow this with a detailed investigation of dynamical switching and dissipation mechanisms.

\begin{table}[H]
\centering
\caption{Main simulation parameters used throughout this work.}
\label{tab:simulation_parameters}
\begin{tabular}{|l|c|}
\hline
\textbf{Quantity} & \textbf{Value} \\
\hline
$\alpha$ & 0.1 \\
$M_s$ & $4.77 \times 10^{5}\ \mathrm{A\,m^{-1}}$ \\
$K_{\mathrm{u}}$ & $11000\ \mathrm{J\,m^{-3}}$ \\
$T$ & $300\ \mathrm{K}$ \\
$N_{\mathrm{part}}$ & 1000 \\
$d$ & $30,\ 50\ \mathrm{nm}$ \\
$\eta$ & $0.00089,\ 0.002,\ 0.044\ \mathrm{Pa\,s}$ \\
$f$ & $0.1,\ 1,\ 10,\ 100\ \mathrm{MHz}$ \\
\hline
\end{tabular}
\end{table}

\section{Easy-axis reorientation under AC fields}
\label{sec:easy_axis_reorientation}

In this section, we characterize the reorientation of the nanoparticle easy axes under an alternating magnetic field. At this stage, we focus not on the heating efficiency itself, but on the orientational dynamics of the particle ensemble, since the time-dependent distribution of easy-axis directions is expected to play a central role in the subsequent magnetic response \cite{serantes2018anisotropic}.

We first examine the transient evolution of the orientational distribution and the characteristic times required to reach a stationary regime. We then analyze the stationary easy-axis configurations and identify a crossover between predominantly perpendicular and parallel/antiparallel orientations. Finally, we investigate how this behaviour depends on particle size, medium viscosity, and excitation frequency. 

\begin{figure*}[!t]
    \centering
    \includegraphics[width=0.8\textwidth]{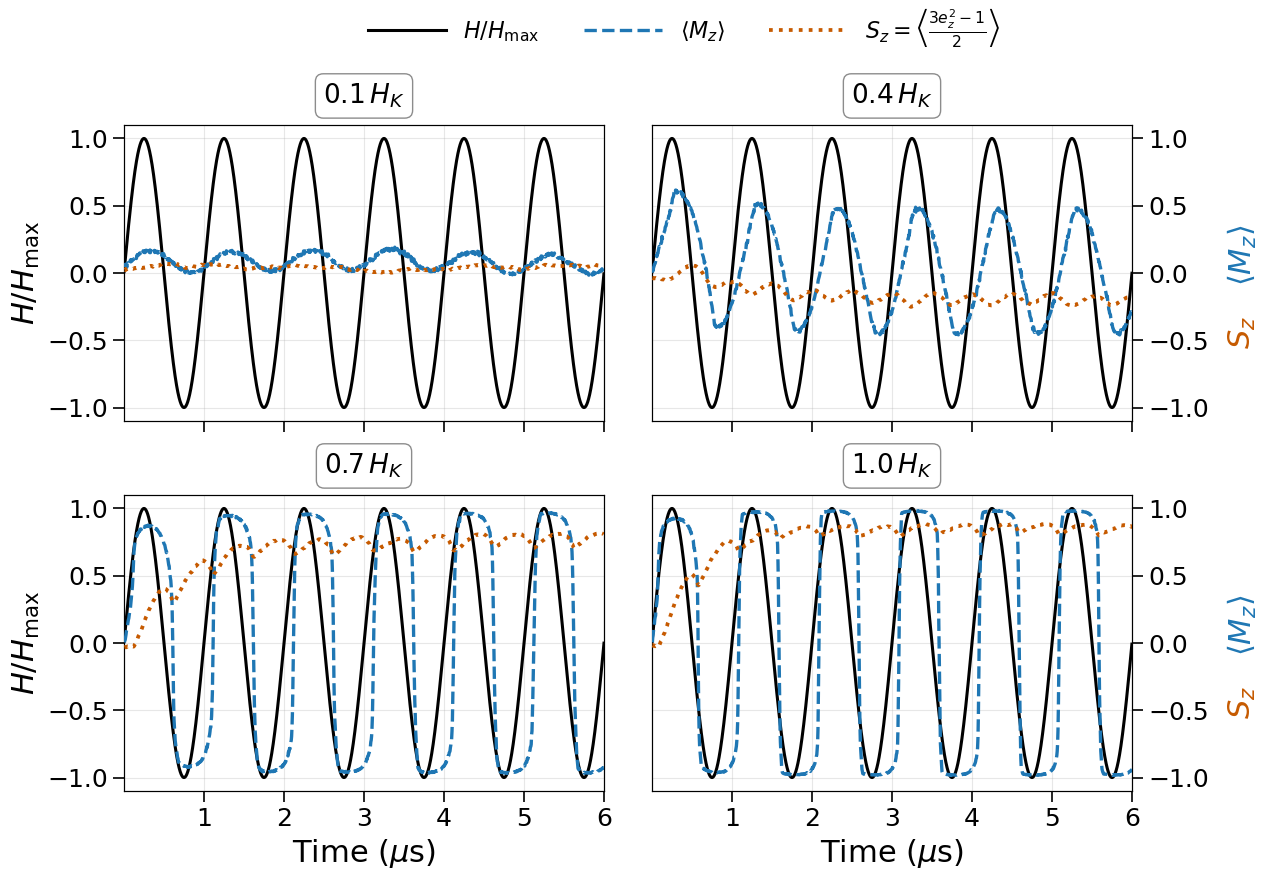}
    \caption{Time evolution of the ensemble-averaged magnetization component along the field direction, $\langle M_z/M_s \rangle$ (blue dashed line), and of the orientational order parameter, $S_z$ (orange dotted line). The normalized applied field $H/H_{max}$ is also shown for reference (black line). Results correspond to particles with diameter $d=30~\mathrm{nm}$  dispersed in water ($\eta=0.00089~\mathrm{Pa\cdot s}$) and driven by an AC field of $f=1~\mathrm{MHz}$. Four representative reduced field amplitudes are shown: $H_{max}/H_k=0.1$, $0.4$, $0.7$, and $1.0$, which correspond to $H_{max}=3.65\times10^3$, $1.46\times10^4$, $2.55\times10^4$, and $3.65\times10^4~\mathrm{A/m}$, respectively.}
    \label{fig:moment_orientation}
\end{figure*}

To characterize the orientational configuration of the ensemble, we evaluate the orientational order parameter $S_z$ \cite{okada2025proposal}, 
\begin{equation}
S_z=\left\langle \frac{3e_z^2-1}{2}\right\rangle,
\label{eq:Sz}
\end{equation}
where $e_z$ is the projection of the easy axis along the field direction. 
The limiting values $S_z=1$ and $S_z=-1/2$ correspond, respectively, to perfectly parallel/antiparallel and perfectly perpendicular easy-axis configurations, while $S_z=0$ corresponds to an isotropic distribution. Because $S_z$ depends on $e_z^2$, it treats parallel and antiparallel easy-axis orientations equivalently.
We also analyze the probability distribution of the easy-axis angle $\theta_K$, defined as the angle between $\hat{\mathbf e}$ and the field direction, as illustrated in Figure \ref{fig:scheme}. This distribution provides a more detailed representation of the orientational state than the scalar order parameter alone.

\subsection{Transient reorientation dynamics}
\label{sec:transient}

As a representative example, we consider an ensemble of MNPs of diameter $d = 30~\mathrm{nm}$ dispersed in water ($\eta = 0.00089~\mathrm{Pa\cdot s}$) and driven by an AC field with frequency $f = 1~\mathrm{MHz}$ at several representative values of  $H_{max}$. 
Figure~\ref{fig:moment_orientation} shows the time dependence of the average magnetization component along the field direction, $\langle M_z /M_s\rangle$, together with the easy-axis orientational order parameter $S_z$. 
The field amplitudes $H_{max}$ are expressed in units of the uniaxial anisotropy field, $H_k=2K_{\mathrm{u}}/(\mu_0 M_s)$, which provides the natural field scale
for describing the magnetic response of uniaxial particles \cite{yanes2007effective,collings2023generalized} and is particularly
relevant for interpreting MFH performance \cite{ruta2015unified,serantes2018anisotropic,Cam2026}.

A clear field-dependent trend is observed in Figure~\ref{fig:moment_orientation}. 
As $H_{max}$ increases, the amplitude of the oscillatory $\langle M_z/M_s\rangle$ response increases and approaches saturation at approximately $H_{max}\sim 0.7H_k$. 
The long-time behaviour of $S_z$ also changes systematically with field amplitude. At low
fields, its negative values indicate a preferentially perpendicular easy-axis configuration, whereas its increase towards positive values at higher fields signals the development of preferential parallel/antiparallel alignment. Because $S_z$ is an ensemble-averaged scalar quantity, we next examine the full angular distribution to characterize this reorientation in greater detail.

\begin{figure*}[!t]
    \centering
    \includegraphics[width=0.95\textwidth]{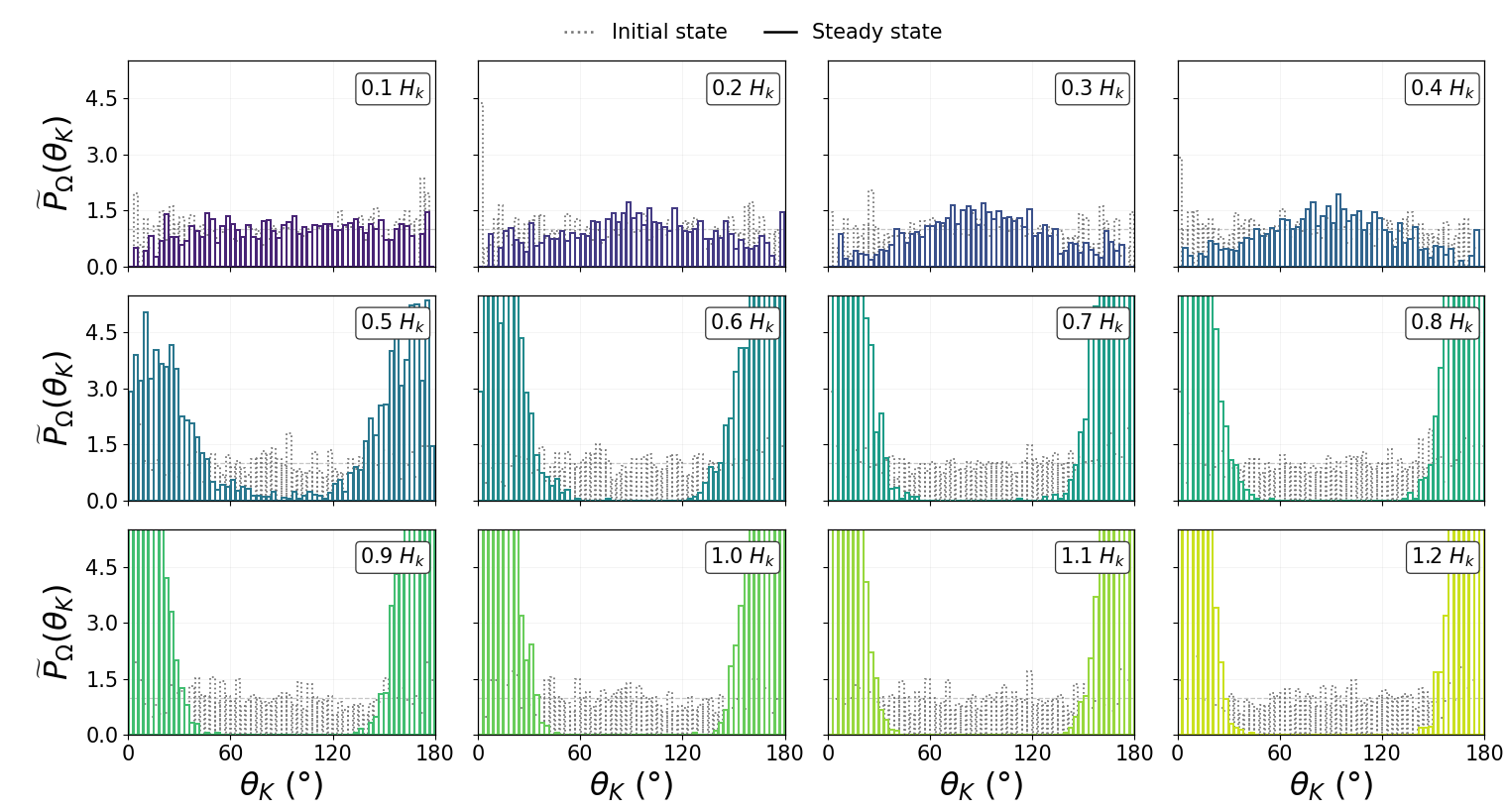}
    \caption{Solid-angle-normalized probability distribution of the easy-axis angle, $\widetilde{P}_\Omega(\theta_K)$, for the different values of $H_{max}/H_k$ considered. 
    The initial isotropic distribution is compared with the distribution obtained in the long-time periodic regime. 
   Results correspond to particles with diameter $d=30~\mathrm{nm}$ dispersed in water and driven by an AC field with frequency $f=1~\mathrm{MHz}$.}
    \label{fig:easy_axis_hist}
\end{figure*}

Figure ~\ref{fig:easy_axis_hist} shows the probability $\widetilde{P}_{\Omega}(\theta_K)$ normalized to the solid angle $\Delta \Omega$ as a function of $\theta_K$. The panels show the evolution as a function of reduced field amplitudes $H_{\max}/H_k$, comparing the probability in the initial random state and in the long-time periodic regime. For each angular bin $[\theta_i,\theta_{i+1}]$, the normalized probability is defined as
\begin{equation}
    \widetilde{P}_{\Omega,i}
    =
    \frac{4\pi N_i}{N\Delta\Omega_i},
    \qquad
    \Delta\Omega_i
    =
    2\pi\left[
    \cos\theta_i-\cos\theta_{i+1}
    \right],
\end{equation}
where $N_i$ is the number of easy axes in the bin and $N$ is the total
number of particles. With this normalization, all distributions integrate to a probability $\int \widetilde{P}_{\Omega} d\Omega=1$. 
Here, $\theta_K$ is the angle between the easy axis and the applied-field direction, as defined in Figure~\ref{fig:scheme}, and is represented over the full interval $[0^\circ,180^\circ]$. Perpendicular orientations correspond to angles near $90^\circ$, whereas parallel and antiparallel orientations correspond to angles near $0^\circ$ and $180^\circ$, respectively.


As shown in Figure ~\ref{fig:easy_axis_hist}, the angular distributions
reveal two distinct orientational regimes. 
At low field amplitudes, increasing $H_{max}$ enhances the statistical weight near $90^\circ$ relative to the initial isotropic distribution, indicating progressively stronger perpendicular ordering. 
This tendency persists up to approximately $H_{max}=0.4H_k$, which marks the onset of the crossover. At higher field
amplitudes, the distribution instead develops increasing statistical weight near $0^\circ$ and $180^\circ$, demonstrating a reorientation towards preferentially parallel or antiparallel configurations.

\begin{figure}[!ptb]
    \centering
    \includegraphics[width=1.0\columnwidth]{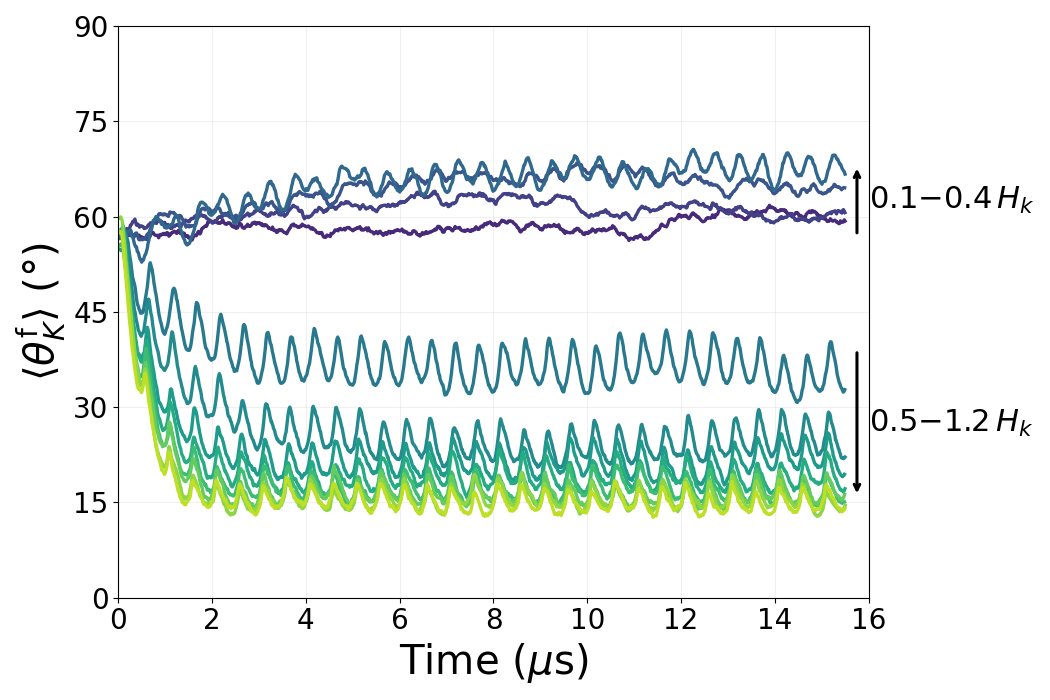}
    \caption{Time evolution of the ensemble-averaged folded easy-axis angle, $\langle \theta_K^{\mathrm{f}} \rangle$. Results correspond to particles with diameter $d=30~\mathrm{nm}$ dispersed in water and driven at $f=1~\mathrm{MHz}$.}
    \label{fig:angle_easy_axis}
\end{figure}

The transient evolution of this reorientation is further illustrated in Figure~\ref{fig:angle_easy_axis} using the folded easy-axis angle $\langle \theta_K^{\mathrm f} \rangle$. For this averaged quantity, angles above $90^\circ$ are mapped onto their supplementary angles below $90^\circ$, so that parallel and antiparallel orientations are treated equivalently and the effective angular range is reduced to the interval $[0^\circ,90^\circ]$. 
Accordingly, $\langle\theta_K^{\mathrm f}
\rangle$ should be regarded as an effective orientational indicator, rather than as 
the mean of the full angular distribution over $[0^\circ,180^\circ]$. 
At low field amplitudes, $\langle\theta_K^{\mathrm f}\rangle$ remains at relatively large values throughout the simulation, consistent with the persistence of the predominantly perpendicular configuration. Once the field amplitude exceeds the crossover region, the folded mean angle decreases rapidly and approaches a much smaller long-time value, corresponding to the parallel/antiparallel regime.

Above the crossover, the time required to reach the long-time orientational regime decreases as $H_{max}$ increases. 
This behaviour is consistent with the larger field-induced excursions of the magnetization, which enhance the anisotropy-mediated torque responsible for particle reorientation over the parameter range considered.
In the representative cases shown here, the main reorientation process occurs over a few microseconds, corresponding to only a few field cycles at $f=1~\mathrm{MHz}$. Thus, increasing the field amplitude modifies not only the preferred long-time orientation of the easy axes but also the rate at which this orientational regime is established.

Figure~\ref{fig:angle_easy_axis} also reveals residual periodic oscillations
around the long-time mean value of $\langle\theta_K^{\mathrm f}\rangle$, particularly in the parallel/antiparallel regime. These oscillations arise from the periodically
driven magnetization dynamics and the resulting time-dependent anisotropy
torque acting on the particle body that induces a small oscillatory motion of the easy axes about the equilibrium average value. 
The long-time state should therefore not be interpreted as a static easy-axis configuration, but as a periodic orientational regime in which the easy axes oscillate around a stable mean orientation.

\subsection{Stationary orientational regimes}
\label{sec:stationary_alignment}

To characterize the final orientational state reached under AC driving, we now analyze the stationary value of the easy-axis angle with respect to the field direction, $\langle\theta_K\rangle_{\mathrm{final}}$, as a function of the reduced field amplitude $H_{max}/H_k$. 
The quantity $\langle\theta_K\rangle_{\mathrm{final}}$ is obtained by averaging the folded easy-axis angle over the particle ensemble and over the final $30~\mu\mathrm{s}$ of the simulation, after the system has reached a periodically stationary state.
In practice, stationarity is identified by comparing the time-averaged orientation in consecutive $10~\mu\mathrm{s}$ windows and requiring the absolute difference between them to remain below $1^\circ$.
%
We also verified that the stationary orientational state is independent of the initial configuration, whether the magnetic moments and easy axes are initialized randomly or preferentially parallel or perpendicular to the applied field.

\begin{figure}[!ptb]
    \centering 
    \includegraphics[width=1.0\columnwidth]{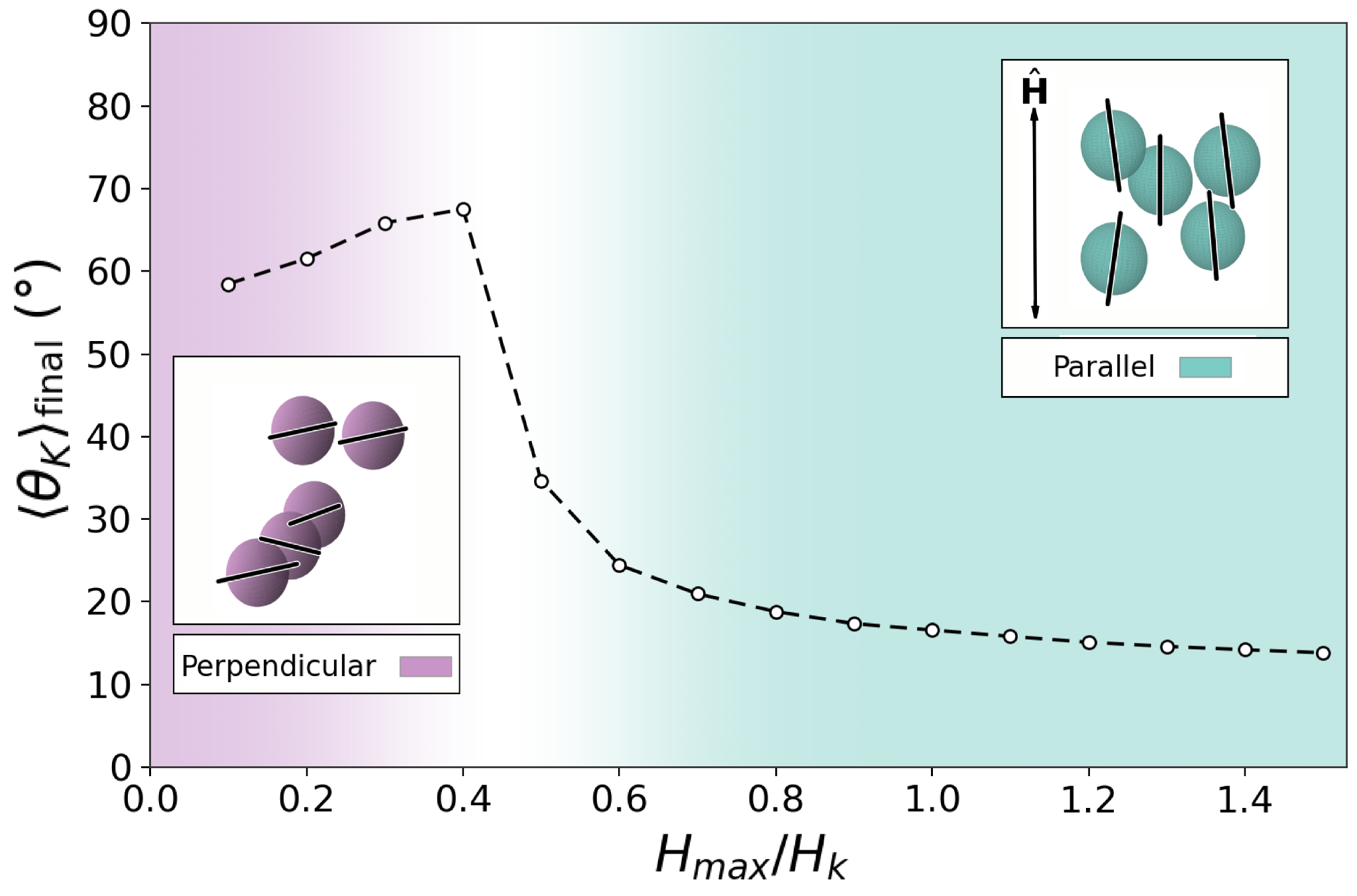}
    \caption{Stationary easy-axis orientation, $\langle \theta_K \rangle_{\mathrm{final}}$, as a function of $H_{max}/H_k$. The insets schematically represent the corresponding easy-axis configurations with respect to the applied field direction.}
\label{fig:theta_final_30nm_water}
\end{figure}

Figure ~\ref{fig:theta_final_30nm_water} shows $\langle \theta_K \rangle_{\mathrm{final}}$ \textit{vs.} $H_{max}/H_k$, under the same conditions as in Section \ref{sec:transient}: $d=30~\mathrm{nm}$ particles in water, subjected to an AC field of $f=1$ MHz. 
The results reveal a clear crossover from a predominantly perpendicular orientational regime at low field amplitudes to a predominantly parallel or antiparallel regime at higher amplitudes.
At small $H_{max}/H_k$, $\langle\theta_K\rangle_{\mathrm{final}}$ remains relatively large, indicating that the easy axes preferentially orient perpendicular to the applied field, as schematically represented in the left inset. 
%
As the field amplitude increases through the crossover region, $\langle\theta_K\rangle_{\mathrm{final}}$ decreases markedly, indicating that the easy axes become preferentially oriented parallel or antiparallel to the field, as illustrated in the right inset.
This behaviour is consistent with previous reports of the field-induced orientational response of anisotropic magnetic nanoparticles under alternating magnetic fields \cite{mamiya2011hyperthermic,ota2021empirical,coene2020simultaneous}.
The principal feature of these data is therefore a crossover between the two orientational regimes over the interval $H_{max}\simeq 0.4$--$0.5H_k$.

\subsection{Dependence on frequency, viscosity, and particle size}
\label{sec:size_viscosity_frequency}

\begin{figure*}[!ptb]
    \centering
    \includegraphics[width=1.0\textwidth]{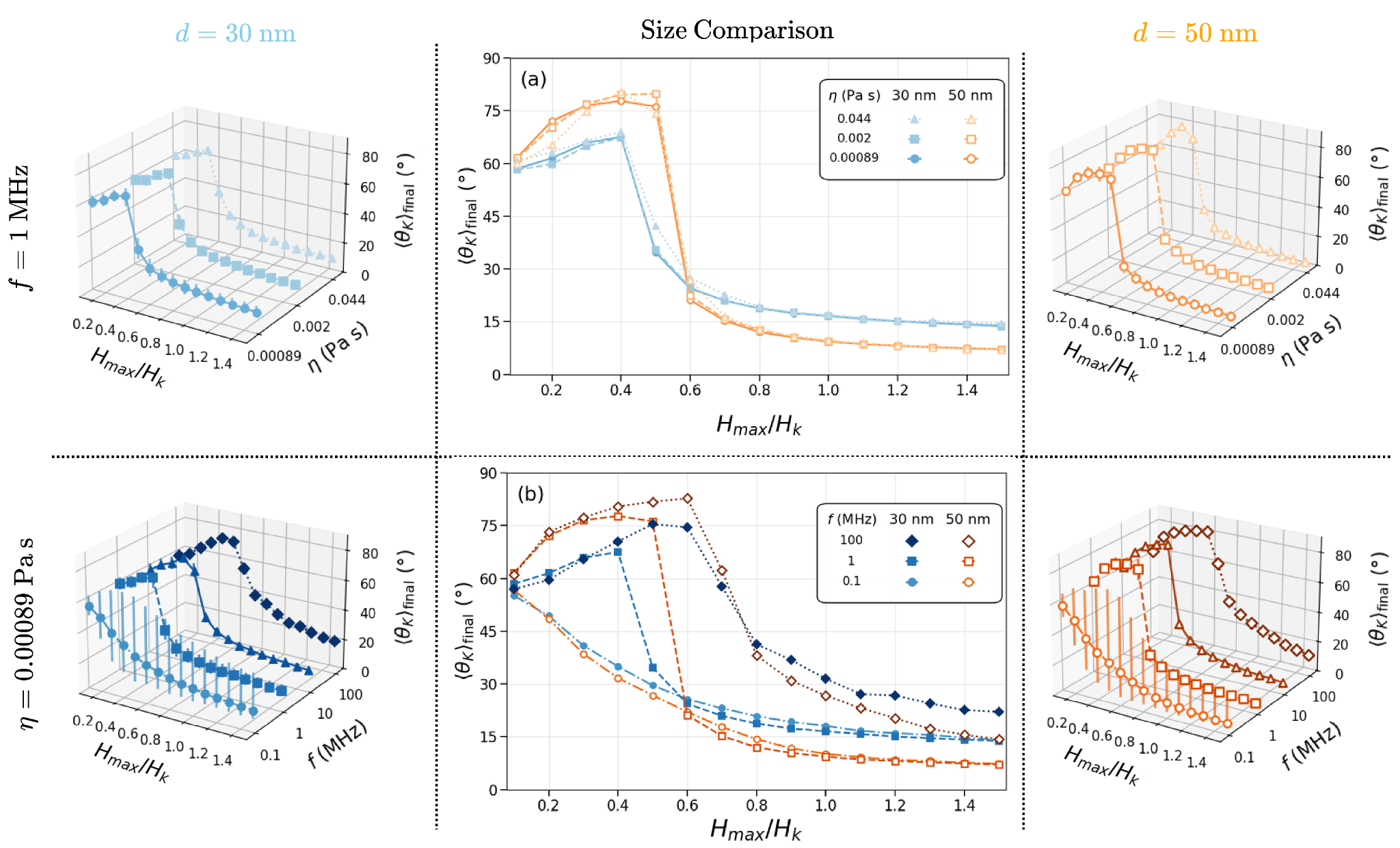}
    \caption{Stationary easy-axis orientation
    $\langle \theta_K \rangle_{\mathrm{final}}$ as a function of $H_{max}/H_k$.
    Top panels: fixed $f=1~\mathrm{MHz}$ and varying viscosity,
    $\eta=0.00089$, $0.002$, and $0.044~\mathrm{Pa\cdot s}$. Bottom panels: fixed $\eta=0.00089~\mathrm{Pa\cdot s}$ and varying frequency, $f=0.1$, $1$, $10$, and $100~\mathrm{MHz}$.
    In each row, the central panel compares the results for $d=30$ and  $50~\mathrm{nm}$, whereas the left and right panels show the corresponding data separately for $d=30$ and $50~\mathrm{nm}$, respectively.
    Angles are folded onto $[0^\circ,90^\circ]$, so that
    parallel and antiparallel orientations are treated equivalently.
     Vertical bars indicate the maximum amplitude of the oscillations in the stationary state, as illustrated in Figure ~\ref{fig:angle_easy_axis}.}
\label{fig:theta_final_size}
\end{figure*}

We now investigate how the crossover from predominantly perpendicular
to predominantly parallel or antiparallel easy-axis orientations
depends on particle size, medium viscosity, and excitation frequency.
To cover a broad range of physical conditions while keeping the
computational cost manageable, we consider four frequencies spanning
three orders of magnitude ($f=0.1$, $1$, $10$, and
$100~\mathrm{MHz}$), three viscosities ($\eta=0.00089$, $0.002$, and
$0.044~\mathrm{Pa\cdot s}$), and two particle diameters ($d=30$ and
$50~\mathrm{nm}$). The lowest viscosity corresponds to water, whereas
$\eta=0.002$ and $0.044~\mathrm{Pa\cdot s}$ represent nanoscale and
macroscopic effective viscosities, respectively, reported for the
HeLa cell cytoplasm \cite{kalwarczyk2011comparative}. For each
parameter combination, we use the same set of 15 reduced field
amplitudes considered in Figure~\ref{fig:theta_final_30nm_water}.
Figure~\ref{fig:theta_final_size} summarizes the results.  The upper row
shows the effect of viscosity at fixed $f=1~\mathrm{MHz}$, whereas the
lower row shows the effect of frequency at fixed
$\eta=0.00089~\mathrm{Pa\cdot s}$. In each row, the central panel
compares both particle sizes, while the left and right panels show the
corresponding results separately for $d=30$ and $50~\mathrm{nm}$,
respectively.

We first consider the role of viscosity, shown in the upper row of Figure~\ref{fig:theta_final_size}. For both particle diameters, the system exhibits the same qualitative crossover identified above: at low field amplitudes, the stationary easy axes remain predominantly perpendicular to the applied field, whereas at larger amplitudes they become preferentially parallel or antiparallel to it. 
Within the resolution of the present field sampling, viscosity does not appreciably shift the location of this crossover.
Its clearest effect is instead to suppress the residual oscillations of the easy-axis orientation in the stationary regime, as indicated by the vertical bars in the side panels. By contrast, particle size influences the crossover field. For $d=30~\mathrm{nm}$, the perpendicular regime begins to break down near $H_{max}\simeq 0.4H_k$, whereas for $d=50~\mathrm{nm}$ it persists to approximately $H_{max}\simeq 0.5H_k$.


We next consider the effect of the excitation frequency, shown in the
lower row of Figure~\ref{fig:theta_final_size}, for particles dispersed
in water.
For $d=30~\mathrm{nm}$, the orientational crossover shifts progressively towards higher field amplitudes, from $\sim 0.4H_k$ at $1~\mathrm{MHz}$ to $\sim 0.5H_k$ at $10~\mathrm{MHz}$ and $\sim0.6H_k$ at $100~\mathrm{MHz}$. 
For $d=50~\mathrm{nm}$, the frequency dependence is weaker: the crossover remains close to $0.5 H_k$ at both $1$ and $10~\mathrm{MHz}$ and shifts to approximately $0.6H_k$ only at $100~\mathrm{MHz}$. A qualitatively different response is observed at $100~\mathrm{kHz}$.
At this frequency, the easy axes evolve monotonically from predominantly perpendicular to predominantly parallel or antiparallel orientations as $H_{max}/H_k$ increases, rather than displaying a narrow crossover. The larger stationary oscillations
observed in this regime are consistent with the longer field period, which allows the easy axis to follow the field-driven magnetization dynamics more closely during each cycle. Overall, the excitation frequency controls both the location and the sharpness of the orientational crossover, as well as the amplitude of the residual
stationary oscillations.

\section{Switching analysis and dissipation mechanisms}
\label{sec:switching}

Having established the easy-axis reorientation under AC field excitation, we now examine the magnetization dynamics associated with the different orientational regimes.
In particular, the crossover from predominantly perpendicular to predominantly parallel or antiparallel easy-axis configurations raises the question of whether it is accompanied by a corresponding change in the reversal dynamics, from non-switching minor-loop trajectories to a regime increasingly dominated by magnetization-switching events.
Characterizing these dynamical regimes is essential for identifying the physical origin of the hysteresis losses and hence their effect on MFH performance.

\subsection{Classification criterion for switching and non-switching trajectories}

To connect the microscopic dynamics with the reorientational regimes, we classify the dynamical trajectories of individual MNPs along the hysteresis loops as \emph{switching} and \emph{non-switching}. 
Switching is detected from the time evolution of the projection of the magnetization direction onto the instantaneous easy-axis direction,
\begin{equation}
s(t)\equiv \hat{\mathbf{m}}(t)\cdot \hat{\mathbf{e}}(t)
= m_x(t)e_x(t)+m_y(t)e_y(t)+m_z(t)e_z(t).
\end{equation}
Thus, $s(t)\simeq+1$ and $s(t)\simeq-1$ correspond to magnetization directions close to the two opposite easy-axis orientations defined by the
instantaneous easy axis. 
Physically, $s(t)$ distinguishes between
cases in which the magnetization lies close to either of the two anisotropy wells associated with the instantaneous easy axis. 
Strictly speaking, under a finite applied field the local energy minima are not generally located exactly at $\hat{\mathbf m}\parallel\pm\hat{\mathbf e}$, because the Zeeman contribution distorts the uniaxial energy landscape and shifts the
corresponding equilibrium directions.
Nevertheless, a sufficiently rapid change of $s(t)$ involving a sign reversal provides a practical indicator of a transition between the two opposite well regions, which we identify as a switching event.

Individual AC periods, $T=1/f$, are analyzed after the stationary
regime has been reached. Each period is divided into two half-loops
corresponding to the two successive branches of the applied field. For
a given half-loop spanning the interval $[t_1,t_2]$, switching is
identified as a sufficiently rapid change of $s(t)$ within a short
event window $\Delta t_{\mathrm{event}}$. Specifically, for every
recorded time $t_i$ satisfying $t_i+\Delta t_{\mathrm{event}}<t_2$, we compute
\begin{equation}
\Delta s(t_i)=s(t_i+\Delta t_{\mathrm{event}})-s(t_i),
\end{equation}
where $s(t_i+\Delta t_{\mathrm{event}})$ is obtained by linear interpolation. A switching event is detected if
\begin{equation}
\max_{t_i\in[t_1,t_2-\Delta t_{\mathrm{event}}]}|\Delta s(t_i)|>\Delta s_{\mathrm{thr}},
\end{equation}
 where $\Delta s_{\mathrm{thr}}$ is a threshold value that we chose based on the observation of the data to ensure that the detected event corresponds to a true reversal between the two opposite well regions, and therefore necessarily involves a sign change of $s(t)$. 
In the present analysis, we use $\Delta t_{\mathrm{event}}=5\times10^{-8}\,\mathrm{s}$ and $\Delta s_{\mathrm{thr}}=1.5$.
Since $s(t)\in[-1,1]$, this amplitude threshold selects rapid variations spanning a substantial fraction of the full range of $s(t)$ and excludes small intra-well fluctuations.

\begin{figure}[H]
    \centering
    \includegraphics[width=1.0\columnwidth]{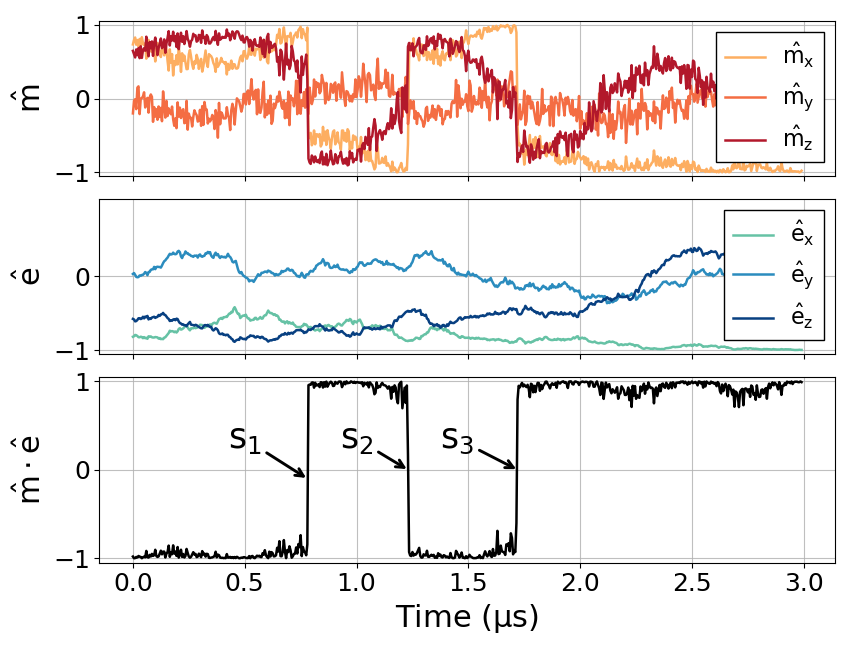}
    \caption{Representative application of the switching-detection criterion to a single particle with a diameter of $d=30~\mathrm{nm}$ dispersed in water and driven at $f=1~\mathrm{MHz}$ with $H_{max}/H_k = 0.4$. Top and middle panels show the time evolution of the components of $\hat{\mathbf{m}}(t)$ and $\hat{\mathbf{e}}(t)$. The bottom panel shows the corresponding projection $s(t)=\hat{\mathbf{m}}(t)\cdot\hat{\mathbf{e}}(t)$. The events labeled $s_1$, $s_2$, and $s_3$ satisfy the switching-detection criterion.}
    \label{fig:switching_repre}
\end{figure}

Each half-loop is classified independently as switching or
non-switching. A complete cycle is considered a full switching cycle
only when a switching event is detected in both half-loops. This
separate analysis of the two field branches distinguishes repeated
field-driven reversal from trajectories containing only an isolated
event. Figure~\ref{fig:switching_repre} illustrates the application of
the criterion.

Figure~\ref{fig:switching_repre} shows rapid changes in the magnetization direction at approximately $0.8$, $1.2$, and $1.7~\mu\mathrm{s}$ (top panel), while the easy-axis direction evolves smoothly across the same events (middle panel). The corresponding rapid sign reversals of $s(t)=\hat{\mathbf m}(t)\cdot\hat{\mathbf e}(t)$ demonstrate that the magnetization moves between the two opposite easy-axis sectors without a simultaneous abrupt rotation of the particle body. These events are therefore identified as internal magnetization reversals rather than changes arising from rigid-body rotation. The bottom panel illustrates how the proposed criterion detects these switching events ($s_1$, $s_2$, and $s_3$) while excluding the smaller, smooth variations of $s(t)$ associated with intra-well oscillations.

\subsection{Field dependence of the switching fraction}

After defining the classification criterion, we quantify how the occurrence of switching events depends on the reduced field amplitude $H_{max}/H_k$. We focus first on MNPs dispersed in water and driven at $1~\mathrm{MHz}$; the corresponding analysis at $100~\mathrm{kHz}$ is presented in Appendix~\ref{appendix_A}. 

\begin{figure*}[!ptb]
    \centering
    \includegraphics[width=1.0\textwidth]{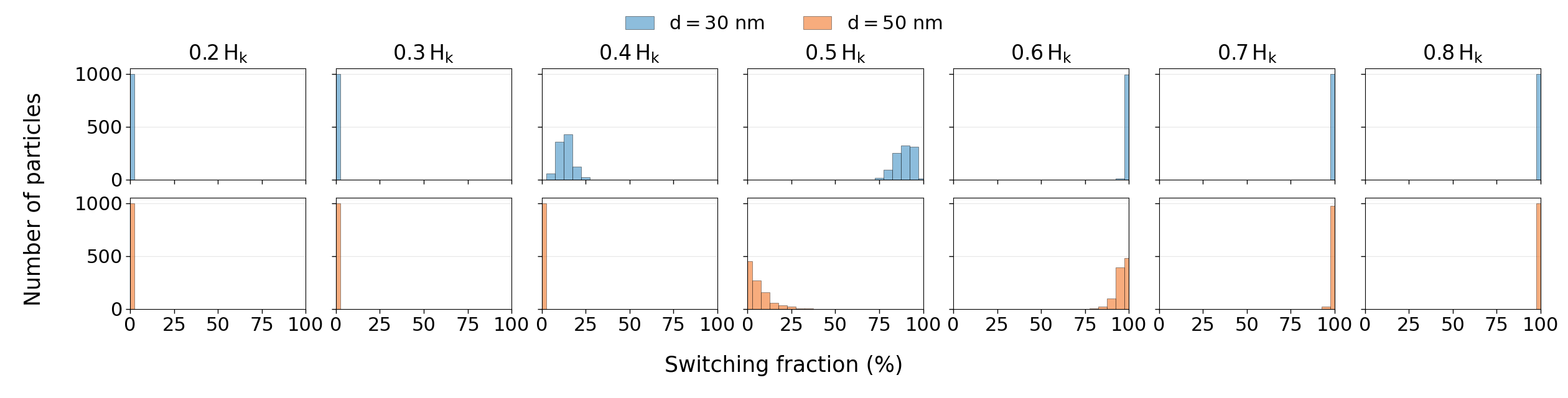}
    \caption{Distribution of the particle-resolved switching fraction
    as a function of the reduced field amplitude $H_{max}/H_k$ for particles with diameters $d=30$ and $50~\mathrm{nm}$ dispersed in water and driven at $f=1~\mathrm{MHz}$. The bar height gives the number of particles with the corresponding switching fraction. The analysis is performed over the final 15 complete field cycles after the stationary regime has been reached.}
    \label{fig:sw_fraction_30}
\end{figure*}

For each value of $H_{max}/H_k$, we analyze only the final 15 complete field cycles after the stationary regime has been reached. The corresponding 30 successive hysteresis half-loops experienced by each MNP are classified as switching or non-switching according to the criterion defined above. 
From this procedure, we obtain the average fraction of half-loops classified as switching for a given $H_{max}/H_k$. The results are shown in Figure~\ref{fig:sw_fraction_30}.

As shown in Figure ~\ref{fig:sw_fraction_30}, no switching events are
detected at low field amplitudes, and all particles exhibit
non-switching minor-loop dynamics. The applied field is therefore
insufficient to drive rapid reversal between the two opposite
easy-axis sectors. Switching events progressively emerge as the field
amplitude increases.
For $d=30~\mathrm{nm}$, switching first appears near $H_{max}\simeq 0.4 H_k$, although it remains intermittent at its onset and occurs in only a small fraction (15-20$\%$) of the analyzed half-loops for most particles. For $d=50~\mathrm{nm}$, the onset shifts to approximately $H_{max}\simeq 0.5H_k$. At this field, nearly half of the particles still exhibit no detected switching, while the remainder display switching with different degrees of intermittency.
At larger field amplitudes, $H_{max}/H_k>0.6$, nearly all analyzed half-loops are classified as switching, showing that the response becomes dominated by rapid internal magnetization reversal, which we associate with
N\'eel-like dynamics.


Notably, the onset fields closely coincide with the orientational crossover identified in the previous section. For particles dispersed in water, the easy-axis reorientation from predominantly perpendicular to predominantly parallel or antiparallel configurations occurs near $0.4H_k$ for $d=30~\mathrm{nm}$ and near $0.5H_k$ for $d=50~\mathrm{nm}$.
This coincidence indicates that the onset of switching and the longitudinal reorientation of the easy axes are closely coupled. Our results are qualitatively consistent with the phenomenology reported by Mamiya and Jeyadevan \cite{mamiya2011hyperthermic}, who associated minor-loop states with blocked magnetization reversal and predominantly transverse easy-axis
orientations, and major-loop states with active reversal and longitudinal easy-axis alignment.

\subsection{Dissipation in switching and non-switching regimes}

\begin{figure}[!ptb]
    \centering
    \includegraphics[width=0.80\columnwidth]{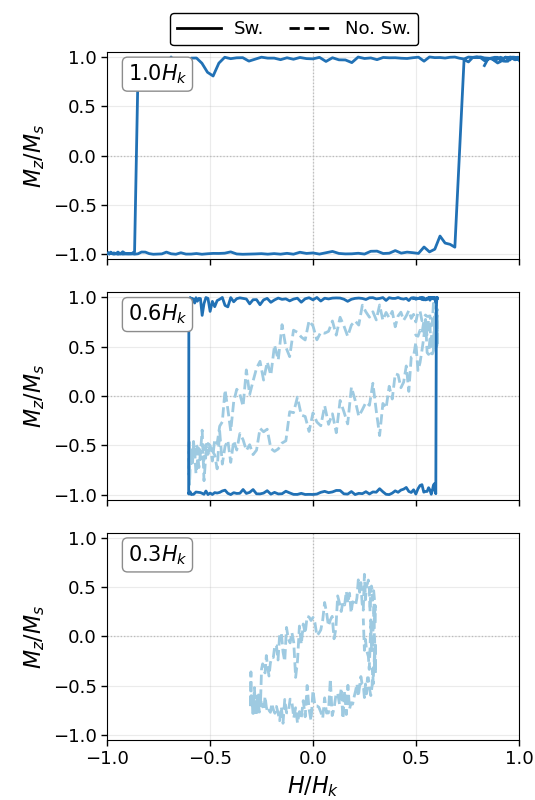}
    \caption{Representative  $M_z(H)$ hysteresis cycles for particles dispersed in water and driven at $f=1~\mathrm{MHz}$.
    Solid lines correspond to switching trajectories and dashed curves to non-switching trajectories. Results are shown for $H_{max}/H_k=1.0$ (top), $0.6$ (middle), and $0.3$ (bottom).}
    \label{fig:cycles_rep}
\end{figure}

Having classified the individual half-loops as switching and non-switching trajectories, we now examine how these dynamical classes are related to energy dissipation.
We first illustrate the two types of response using representative hysteresis cycles in Figure~\ref{fig:cycles_rep}.
We then compare their associated mean loop areas. The energy dissipated per cycle and per unit magnetic-material volume is obtained from the hysteresis-loop area, $A=\mu_0\oint M_z\,dH$. The corresponding specific loss power is
$\mathrm{SLP}=fA/\rho$, where $\rho$ is the mass density of the magnetic material.

Figure~\ref{fig:cycles_rep} shows representative hysteresis cycles for a single particle at different field amplitudes. 
The switching trajectories display comparatively square loops, reflecting rapid internal magnetization reversal between opposite easy-axis sectors. 
By contrast, in the non-switching trajectories (dashed lines) the magnetization remains within the same easy-axis sector and evolves smoothly together with the rotating particle body. 
The examples reflect the statistics reported in Figure~\ref{fig:sw_fraction_30}: switching trajectories dominate at
large fields, non-switching trajectories dominate at small fields (top and bottom panels, respectively), and both types coexist at intermediate amplitudes (middle panel).

\begin{figure*}[!ptb]
    \centering
    \includegraphics[width=0.8\textwidth]{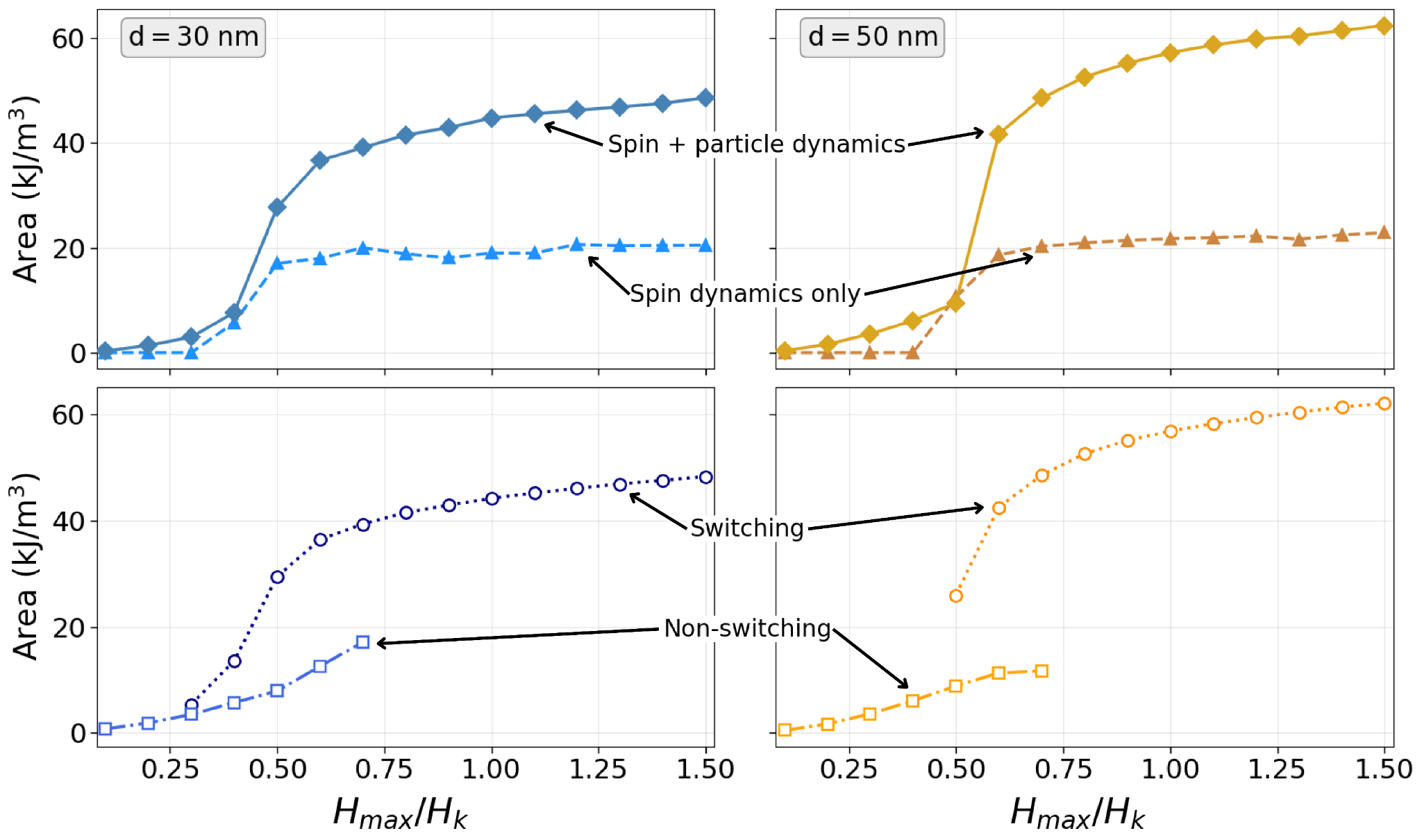}
    \caption{Top panels: ensemble-averaged hysteresis loop area as a function of $H_{max}/H_k$ for the hybrid spin--particle dynamics model of the present work, and for the reference case of mechanically fixed particles with spin dynamics only.
    Bottom panels: separate contributions of the switching and non-switching processes to the loop areas of the spin+particle dynamics cases shown in the top panels. Results are shown for $d=30~\mathrm{nm}$ (left) and $d=50~\mathrm{nm}$ (right), for particles dispersed in water and driven at $f=1~\mathrm{MHz}$. Averages are evaluated over the final 15 field cycles after the stationary regime has been reached.}
    \label{fig:areas_30}
\end{figure*}

We next examine how the two dynamical classes are related to energy dissipation. For each field amplitude, the loop areas are calculated over the final 15 field cycles after the stationary regime has been reached. We first quantify the overall effect of particle rotation by comparing the ensemble-averaged loop area of the coupled spin--particle model with that of a reference model in which the particle orientation is fixed and only the magnetization dynamics is retained. The results are shown in the upper panels of Figure~\ref{fig:areas_30}.

For both particle sizes, allowing physical particle rotation substantially increases the hysteresis losses. For $d=30~\mathrm{nm}$, the coupled and fixed-particle models exhibit a similar onset of dissipation, but their responses separate markedly above approximately $0.4H_k$. At high fields, the loop area approaches $48$--$50~\mathrm{kJ\,m^{-3}}$ in the coupled model, compared with approximately $20~\mathrm{kJ\,m^{-3}}$ for the fixed-particle model. A similar but more pronounced enhancement is observed for $d=50~\mathrm{nm}$: the coupled-model area reaches approximately
$60~\mathrm{kJ\,m^{-3}}$, whereas the spin-only response saturates near $22~\mathrm{kJ\,m^{-3}}$. 
Smaller sizes (not shown) show smaller losses but with an onset of dissipation at smaller field amplitudes, as reported elsewhere \cite{ovejero2021selective}.
Thus, particle rotation does not merely add a low-field dissipative channel, but substantially modifies the total magnetic response over the full field range above the crossover. In addition, whereas the spin-only losses approximately saturate above $H_{max}\simeq0.5$--$0.6H_k$, the losses of the coupled model continue
to increase gradually.


To determine how the dynamical classification is reflected in the losses, the lower panels of Figure~\ref{fig:areas_30} show the conditional mean loop areas of the switching and non-switching populations. At each field amplitude, the switching average is calculated only over trajectories classified as switching, and the non-switching average only over trajectories classified as non-switching. Consequently, the absence of a point indicates that no trajectory of the corresponding class is present at that field amplitude; it does not indicate zero loop area.

Whenever both populations coexist, switching trajectories exhibit larger mean loop areas than non-switching trajectories. For $d=30~\mathrm{nm}$, switching trajectories first appear near $H_{max}\simeq0.4H_k$, and their mean area increases rapidly before approaching the total coupled-model area at high fields. The non-switching trajectories persist up to approximately $0.7H_k$ and have smaller characteristic areas. For $d=50~\mathrm{nm}$, switching first appears near $0.5H_k$, already with a relatively large mean area, whereas the non-switching population again disappears above the crossover region.

We note that the mean loop areas shown in the lower panels of Figure~\ref{fig:areas_30} are conditional on the dynamical class and are not weighted by the frequency with which each class occurs.
They must therefore be interpreted together with the switching fractions in Figure~\ref{fig:sw_fraction_30}. Accordingly, the losses are dominated by the non-switching rotational response at low fields and by N\'eel-like reversal at high fields. In the intermediate coexistence region, the contribution of each class depends on both its characteristic loop area and its frequency of occurrence.
%
We therefore refer to the low-field non-switching and high-field switching responses as Brownian-like and N\'eel-like, respectively.  This terminology describes their predominant dynamical character and does not imply an exact separation of the total losses into purely viscous and purely magnetic damping contributions.
The larger N\'eel-like losses and the higher field amplitudes required to activate them are consistent with previous studies \cite{mamiya2011hyperthermic,serantes2018anisotropic}. A more detailed illustration of the non-switching rotational response is provided in Appendix~\ref{appendix_B}.

Applying the same dynamical classification at the lower excitation
frequency reveals a different balance between switching and
non-switching responses. The corresponding particle-resolved
distributions of switching events at $f=100$ kHz are provided in
Figure~\ref{fig:SI_switching_100kHz} in Appendix~\ref{appendix_A}.
Figure~\ref{fig:areas_100kHz} shows the conditional mean hysteresis-loop
areas of the switching and non-switching populations at $f=100$ kHz.
As in the lower panels of Figure~\ref{fig:areas_30}, the averages
are evaluated separately within each dynamical class and must therefore
be interpreted together with the corresponding switching fractions
reported in Appendix~\ref{appendix_A}.

\begin{figure}[!t]
    \centering
    \includegraphics[width=\columnwidth]{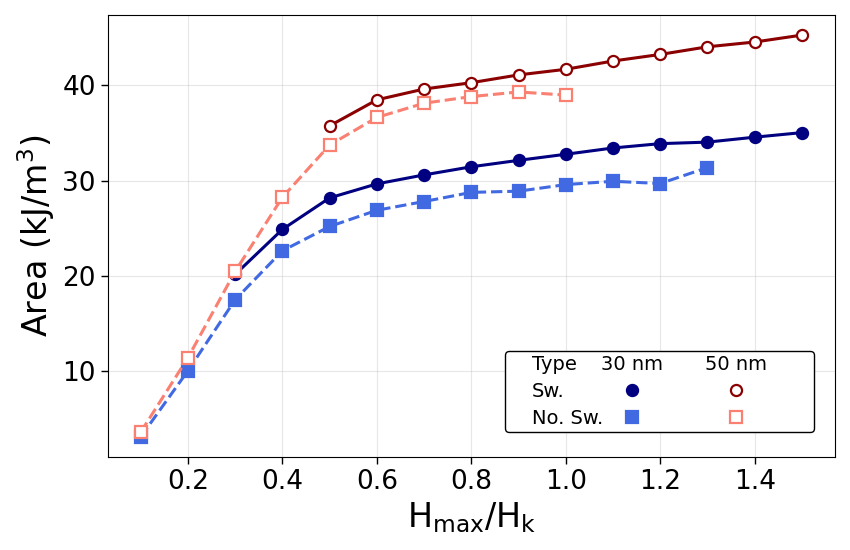}
    \caption{Conditional mean hysteresis-loop area as a function of
    $H_{\max}/H_k$ for trajectories classified as switching and
    non-switching, for particles with diameters $d=30$ and $50$ nm
    dispersed in water and driven at $f=100$ kHz. At each field
    amplitude, the averages are calculated separately over the
    trajectories belonging to the indicated dynamical class. Missing
    points indicate that no trajectory of the corresponding class was
    observed at that field amplitude.}
    \label{fig:areas_100kHz}
\end{figure}
Compared with the results obtained at $f=1$ MHz, the separation between
the two populations is less pronounced at $100$ kHz. Nevertheless, the maximum energy dissipated per cycle decreases only moderately, from approximately $60~\mathrm{kJ\,m^{-3}}$ at $1~\mathrm{MHz}$ to $40~\mathrm{kJ\,m^{-3}}$ at $100~\mathrm{kHz}$ for $H_{max}/H_k=1.5$. 
The corresponding SLP decreases much more strongly because it is proportional to the excitation frequency. Thus, comparable loop areas per cycle do not necessarily imply comparable power dissipation.
The frequency dependence is also apparent in the orientational dynamics: despite the relatively similar loop areas per cycle, the amplitudes of the stationary easy-axis oscillations differ markedly between the two frequencies, as shown in the lower panels of Figure~\ref{fig:theta_final_size}. 
This difference may become particularly relevant for nonspherical particles, for which anisotropic rotational friction and shape anisotropy could modify both the
orientational dynamics and the balance between switching and non-switching losses.

\section{Conclusions}

In this work, we have built upon established coupled magnetic and rotational models to investigate the reorientation of the easy axes of magnetic nanoparticles under AC magnetic fields and its relation to energy dissipation.
Particular attention was paid to the effects of particle size, medium viscosity, and excitation frequency on the orientational dynamics and hysteresis loop losses.

The simulations reveal a field-driven crossover from predominantly perpendicular to predominantly parallel or antiparallel easy-axis configurations. This crossover generally occurs around $H_{max}\simeq0.5H_k$, although its precise location  depends on particle size and excitation frequency. Increasing the frequency shifts the crossover towards higher field amplitudes and also modifies its sharpness: at $100~\mathrm{kHz}$, the orientational evolution with field is considerably more gradual than at higher frequencies.
Within the range investigated, viscosity does not appreciably alter the final mean
orientation or the location of the crossover. Its effects are primarily dynamical, including the suppression of the residual easy-axis oscillations in the stationary regime and an increase in the transient reorientation time.

We then connected these orientational regimes with the microscopic dissipation mechanisms by classifying the analyzed half-loops as switching or non-switching trajectories.
Switching trajectories involve rapid internal magnetization reversal and are therefore associated with N\'eel-like dynamics. In non-switching trajectories, the magnetization remains within the same easy-axis sector and evolves predominantly with the rotating particle, leading to Brownian-like dynamics. 
The onset of switching closely coincides with the orientational crossover towards
parallel or antiparallel easy-axis configurations. For particles
dispersed in water and driven at $f=1~\mathrm{MHz}$, both changes occur
near $0.4H_k$ for $d=30~\mathrm{nm}$ and near $0.5H_k$ for
$d=50~\mathrm{nm}$.

Overall, the present results provide a mechanistic picture of the close coupling between easy-axis reorientation, magnetization switching, and heat generation in the
non-interacting single-particle limit.

Future work should extend the present framework to include interparticle interactions and the associated collective structural evolution, including chain formation and aggregation~\cite{fernandez2024reversible,okada2025proposal}. Further refinements should also relax both the assumption of an ideal spherical particle geometry and the restriction to a purely uniaxial magnetic anisotropy~\cite{failde2024understanding}.
The spherical approximation was adopted here to enable direct comparison with established models and to isolate the coupling between magnetization reversal and particle rotation.

\section*{Author Contributions}

Conceptualization: D.S.;
Data curation: I.L.V.;
Formal analysis:  I.L.V., D.S., Ò.I., R.W.C.; 
Funding acquisition: D.S.;
Investigation:  I.L.V., D.S., Ò.I.;
Methodology:  I.L.V., D.S., Ò.I., R.W.C.;
Project administration: D.S., Ò.I.;
Software: I.L.V.;
Supervision: D.S., Ò.I., R.W.C.;
Validation: I.L.V., D.S., Ò.I.;
Visualization: I.L.V.;
Writing – original draft: I.L.V.;
Writing – review \& editing:  I.L.V., S.U.H., K.O., S.R., R.W.C., D.S., Ò.I.

\section*{Conflicts of interest}

The authors declare no conflict of interest.

\section*{Data availability}

The data that support the findings of this study are available
from the corresponding author upon reasonable request.

\appendix

\titleformat{\section}
  {\sffamily\Large}
  {Appendix \thesection:}
  {0.3em}
  {}

\section{Switching analysis at $f = 100$ kHz} \label{appendix_A}

To complement the switching analysis presented in the main text, we
consider particles dispersed in water and driven by an AC magnetic
field with frequency $f=100$ kHz. The same switching-detection
criterion and averaging protocol described in Section~\ref{sec:switching} of the main text are applied. For each value of $H_{\max}/H_k$, the analysis is
performed over the final 15 complete AC cycles after the stationary
regime has been reached.

\begin{figure}[H]
    \centering
    \includegraphics[width=1.0\columnwidth]{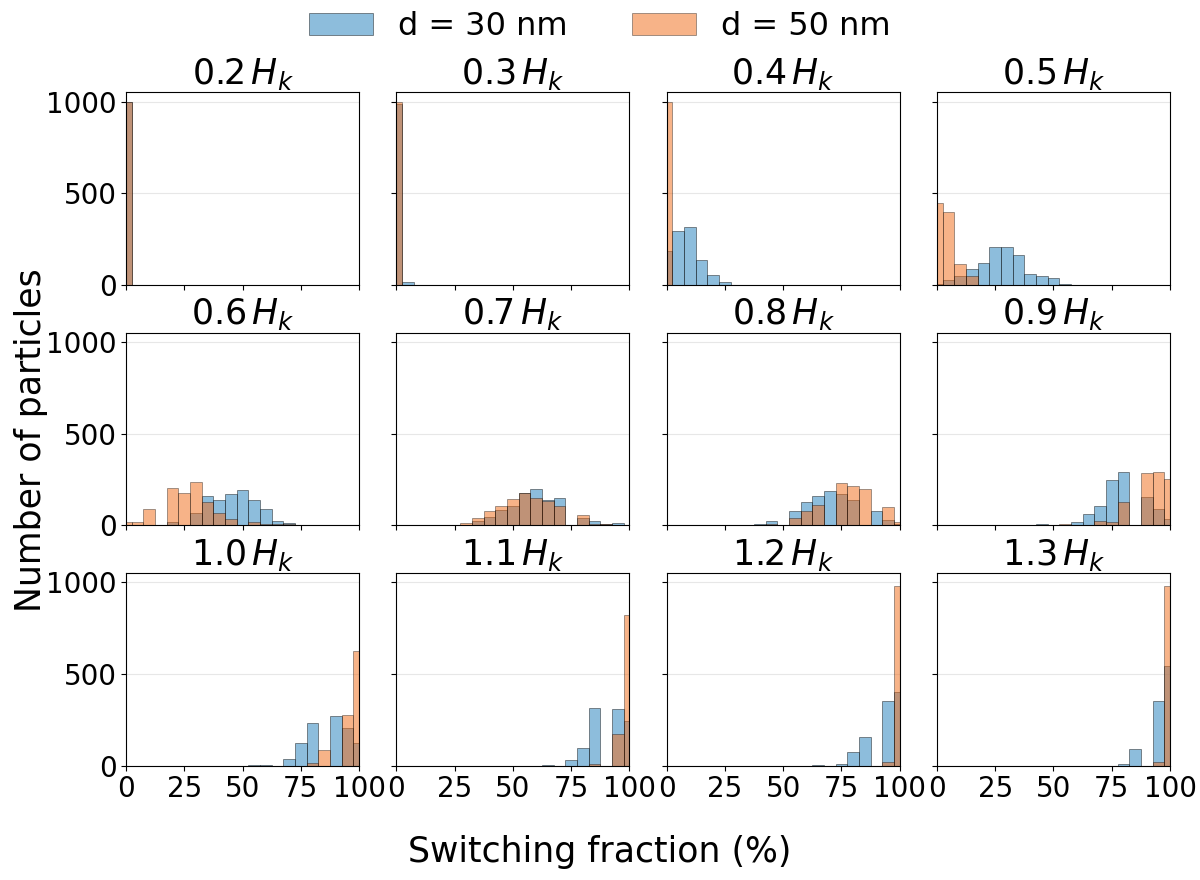}
    \caption{Particle-resolved distribution of the switching fraction
    as a function of $H_{\max}/H_k$ for particles with diameters
    $d=30$ and $50$ nm dispersed in water and driven at
    $f=100$ kHz. The horizontal coordinate in each panel gives the
    percentage of half-loops classified as switching for each particle,
    while the bar height gives the number of particles with the
    corresponding switching fraction. The classification is performed
    over the final 15 complete AC cycles after the stationary regime
    has been reached.}
    \label{fig:SI_switching_100kHz}
\end{figure}

Figure~\ref{fig:SI_switching_100kHz} shows the particle-resolved
switching-fraction distributions for $d=30$ and $50$ nm. In contrast
to the relatively sharp transition observed at $f=1$ MHz, no narrow
crossover between non-switching and switching regimes is found at
$100$ kHz. Instead, the switching fraction increases progressively
over a broad range of field amplitudes.

The first switching events occur at lower values of
$H_{\max}/H_k$ for the $30$ nm particles than for the $50$ nm
particles, indicating that magnetization reversal becomes active at a
smaller reduced field for the smaller particle size. This behaviour is
consistent with the stationary easy-axis orientations reported in the
main text: at $100$ kHz, the reorientation from predominantly
perpendicular to predominantly parallel or antiparallel configurations
also occurs gradually with increasing field amplitude rather than
through a narrow crossover.

At intermediate field amplitudes, the broad distributions demonstrate
that particles with different switching fractions coexist within the
same ensemble. Non-switching and intermittently switching trajectories
therefore remain present over a substantially broader field range than
at $1$ MHz. The conditional mean loop areas associated with these two
dynamical populations are presented and discussed in the main text.

\section{Brownian-like rotational response} \label{appendix_B}

We clarify here why the non-switching trajectories discussed in the main text are referred to as \emph{Brownian-like rotational responses}. 
This terminology is intended to describe their dominant dynamical character: in the absence of reversal between opposite easy-axis sectors, the magnetization evolves approximately together with the rotating particle body.


\begin{figure}[H]
    \centering
    \includegraphics[width=1.0\columnwidth]{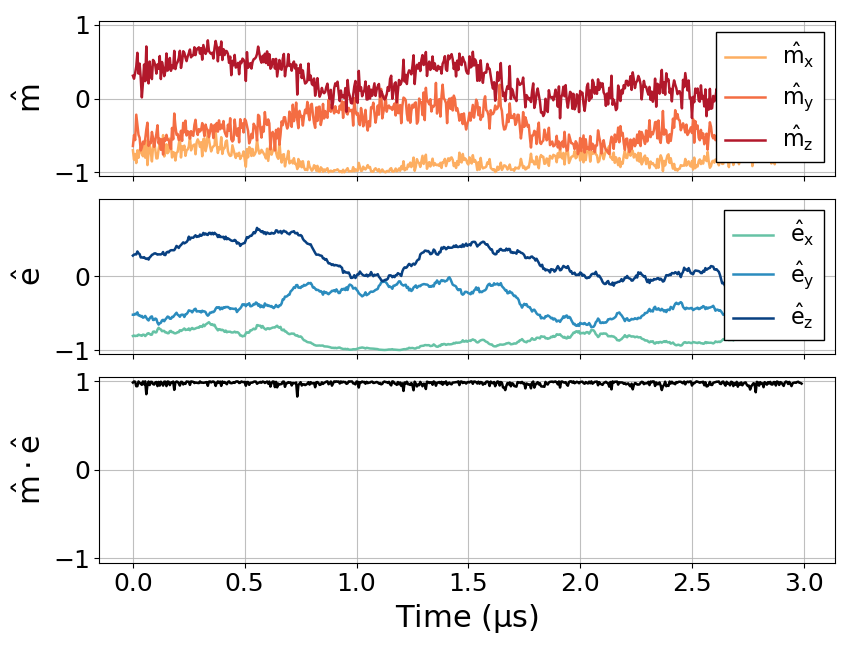}
    \caption{Representative non-switching trajectory illustrating the Brownian-like rotational response for $H_{max}/H_k$ = 0.2. Top: components of the magnetization vector $\hat{\mathbf{m}}$. Middle: components of the easy-axis unit vector $\hat{\mathbf{e}}$. Bottom: time evolution of the projection $\hat{\mathbf{m}}\cdot\hat{\mathbf{e}}$. }
    \label{fig:brownian_response_components}
\end{figure}

\begin{figure}[H]
    \centering
    \includegraphics[width=1.0\columnwidth]{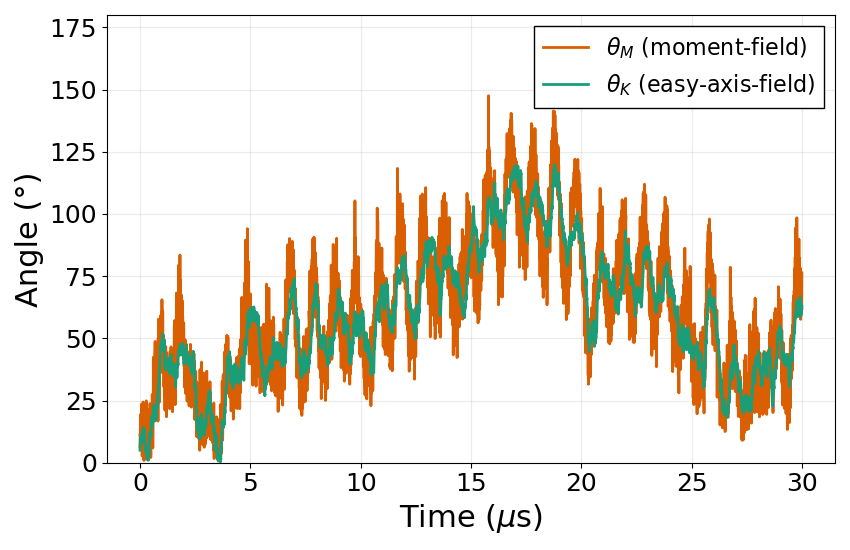}
    \caption{Time evolution of the angle between the magnetic moment and the applied field, $\theta_M$ (orange), and the angle between the easy axis and the field, $\theta_K$ (green), for the same non-switching trajectory shown in Figure ~\ref{fig:brownian_response_components}.}
    \label{fig:brownian_response_angles}
\end{figure}

As illustrated in Figure ~\ref{fig:brownian_response_components} for a given particle subjected to $H_{max}/H_k$ = 0.2, when no switching event takes place the projection $\hat{\mathbf{m}}\cdot\hat{\mathbf{e}}$ remains positive and close to unity throughout the cycle. This indicates that the magnetic moment stays largely confined to the same anisotropy well and remains approximately locked to the easy axis. In this regime, the evolution of $\hat{\mathbf{m}}$ closely follows the reorientation of $\hat{\mathbf{e}}$, so that the magnetic response is dominated by the joint rotation of the particle body and its easy axis rather than by irreversible N\'eel-like reversal between anisotropy minima.

This interpretation is further supported by Figure ~\ref{fig:brownian_response_angles}, where the angle between the magnetic moment and the applied field, $\theta_M$, and the angle between the easy axis and the applied field, $\theta_K$, are shown as a function of time. In the non-switching regime, both quantities exhibit closely correlated temporal variations. In particular, under minor-loop conditions the easy axis remains close to a nearly perpendicular configuration with respect to the field, while the magnetization follows the field-driven oscillation without undergoing full reversal. 
The close correspondence between
$\theta_M(t)$ and $\theta_K(t)$ therefore supports the interpretation
of these trajectories as rotational or Brownian-like.

\section*{Acknowledgements}

The authors gratefully acknowledge Prof. Akira Satoh (Professor Emeritus, Akita Prefectural University) for his valuable scientific advice and support, which were essential to the development of the model presented in this work. This work was supported by the Spanish Ministry of Science, Innovation and Universities (MICIU) and the State Research Agency (AEI, 10.13039/501100011033) through Grants No. PID2019-109514RJ-I00, No. PID2024-157172NB-I00, No. PID2025-171111OB-I00 and No. CNS2024-154574, and through a Ramón y Cajal Fellowship (No. RYC2020-029822-I) to D.S.; by the European Regional Development Fund (ERDF/EU) through Grant No. PID2024-157172NB-I00; by the Generalitat de Catalunya through Grant No. 2021SGR0032; and by the Xunta de Galicia through Grants No. ED431F 2022/005 and No. ED431B 2023/055. We acknowledge the Centro de Supercomputación de Galicia (CESGA) for providing computing resources. This work was supported in part by JSPS KAKENHI Grant Number JP26K17308.



\balance

\renewcommand\refname{References}

\bibliography{master_bibliography_cleaned_updated} 

@article{behbahani2022micromagnetic,
  author = {Behbahani, Razyeh and Plumer, Martin L and Saika-Voivod, Ivan},
  journal = {Phys. Rev. Appl.},
  number = {3},
  pages = {034034},
  publisher = {APS},
  title = {Micromagnetic simulations of clusters of nanoparticles with internal structure: Application to magnetic hyperthermia},
  volume = {18},
  year = {2022},
  doi = {10.1103/PhysRevApplied.18.034034}
}

@Article{Beola2020,
  author = {Beola, Lilianne and As{\'\i}n, Laura and Roma-Rodrigues, Catarina and Fern{\'a}ndez-Afonso, Yilian and Fratila, Raluca M and Serantes, David and Ruta, Sergiu and Chantrell, Roy W and Fernandes, Alexandra R and Baptista, Pedro V and others},
  fjournal = {ACS Applied Materials & Interfaces},
  journal = {ACS Appl. Mater. Interfaces},
  number = {39},
  pages = {43474--43487},
  publisher = {ACS Publications},
  title = {The intracellular number of magnetic nanoparticles modulates the apoptotic death pathway after magnetic hyperthermia treatment},
  volume = {12},
  year = {2020},
  doi = {10.1021/acsami.0c12900}
}

@article{cabrera2017unraveling,
  author = {Cabrera, David and Lak, Aidin and Yoshida, T and Materia, Maria Elena and Ortega, D and Ludwig, Franz and Guardia, Pablo and Sathya, A and Pellegrino, Teresa and Teran, Francisco J},
  journal = {Nanoscale},
  number = {16},
  pages = {5094--5101},
  title = {Unraveling viscosity effects on the hysteresis losses of magnetic nanocubes},
  volume = {9},
  year = {2017},
  doi = {10.1039/C7NR00810D}
}

@article{coene2020simultaneous,
  author = {Coene, Annelies and Leliaert, Jonathan},
  journal = {Sensors},
  number = {14},
  pages = {3882},
  publisher = {MDPI},
  title = {Simultaneous coercivity and size determination of magnetic nanoparticles},
  volume = {20},
  year = {2020},
  doi = {10.3390/s20143882}
}

@article{collings2023generalized,
  author = {Collings, Jack B and Rama-Eiroa, Ricardo and Otxoa, Rub{\'e}n M and Evans, Richard FL and Chantrell, Roy W},
  journal = {Phys. Rev. B},
  number = {6},
  pages = {064413},
  publisher = {APS},
  title = {Generalized form of the magnetic anisotropy field in micromagnetic and atomistic spin models},
  volume = {107},
  year = {2023},
  doi = {10.1103/PhysRevB.107.064413}
}

@article{del2022magnetogenetics,
  author = {Del Sol-Fern{\'a}ndez, Susel and Mart{\'\i}nez-Vicente, Pablo and Gomoll{\'o}n-Zueco, Pilar and Castro-Hinojosa, Christian and Guti{\'e}rrez, Luc{\'\i}a and Fratila, Raluca M and Moros, Mar{\'\i}a},
  doi = {10.1039/D1NR06303K},
  journal = {Nanoscale},
  number = {6},
  pages = {2091--2118},
  publisher = {Royal Society of Chemistry},
  title = {Magnetogenetics: remote activation of cellular functions triggered by magnetic switches},
  volume = {14},
  year = {2022}
}

@article{di2014magnetic,
  author = {Di Corato, Riccardo and Espinosa, Ana and Lartigue, Lenaic and Tharaud, Mickael and Chat, Sophie and Pellegrino, Teresa and M{\'e}nager, Christine and Gazeau, Florence and Wilhelm, Claire},
  journal = {Biomaterials},
  number = {24},
  pages = {6400--6411},
  publisher = {Elsevier},
  title = {Magnetic hyperthermia efficiency in the cellular environment for different nanoparticle designs},
  volume = {35},
  year = {2014},
  doi = {10.1016/j.biomaterials.2014.04.036}
}

@article{dormand1980family,
  author = {Dormand, John R and Prince, Peter J},
  journal = {J. Comput. Appl. Math.},
  number = {1},
  pages = {19--26},
  title = {A family of embedded Runge-Kutta formulae},
  doi = {10.1016/0771-050X(80)90013-3},
  volume = {6},
  year = {1980}
}

@article{durhuus2024conservation,
  author = {Durhuus, Frederik L and Beleggia, Marco and Frandsen, Cathrine},
  doi = {10.1103/PhysRevB.109.054421},
  journal = {Phys. Rev. B},
  number = {5},
  pages = {054421},
  publisher = {APS},
  title = {Conservation laws for interacting magnetic nanoparticles at finite temperature},
  volume = {109},
  year = {2024}
}

@article{failde2024understanding,
  author = {Fa{\'\i}lde, Daniel and Ocampo-Zalvide, Victor and Serantes, David and Iglesias, {\`O}scar},
  journal = {Nanoscale},
  number = {30},
  pages = {14319--14329},
  publisher = {Royal Society of Chemistry},
  title = {Understanding magnetic hyperthermia performance within the “Brezovich criterion”: beyond the uniaxial anisotropy description},
  volume = {16},
  year = {2024},
  doi = {10.1039/D4NR02045F}
}

@article{fernandez2024reversible,
  author = {Fern{\'a}ndez-Afonso, Yilian and Ruta, Sergiu and P{\'a}ez-Rodr{\'\i}guez, Amira and van Zanten, Thomas S and Gleadhall, Sian and Fratila, Raluca M and Moros, Mar{\'\i}a and Morales, Maria del Puerto and Satoh, Akira and Chantrell, Roy W and others},
  journal = {Adv. Funct. Mater.},
  number = {40},
  pages = {2405334},
  publisher = {Wiley Online Library},
  title = {Reversible alignment of nanoparticles and intracellular vesicles during magnetic hyperthermia experiments},
  volume = {34},
  year = {2024},
  doi = {10.1002/adfm.202405334}
}

@article{gavilan2025magnetic,
  author = {Gavil{\'a}n, Helena and Gallo-Cordova, Alvaro and Chediak, Maura Lisett R{\'a}bade and Rodr{\'\i}guez, Amira P{\'a}ez and del Puerto Morales, Maria and Guti{\'e}rrez, Luc{\'\i}a},
  doi = {10.1039/d5nr03329b},
  issue = "48",
  journal = {Nanoscale},
  pages = "27734-27761",
  title = {Magnetic hyperthermia in focus: emerging non-cancer applications of magnetic nanoparticles},
  volume = {17},
  year = {2025}
}

@book{hairer1993solving,
  author = {Hairer, Ernst and Wanner, Gerhard and N{\o}rsett, Syvert P},
  publisher = {Springer},
  title = {Solving ordinary differential equations I: Nonstiff problems},
  year = {1993},
  doi = {10.1007/978-3-540-78862-1}
}

@article{healy2022clinical,
  author = {Healy, Sean and Bakuzis, Andris F and Goodwill, Patrick W and Attaluri, Anilchandra and Bulte, Jeff WM and Ivkov, Robert},
  fjournal = {Wiley Interdisciplinary Reviews: Nanomedicine and Nanobiotechnology},
  journal = {WIREs Nanomed. Nanobiotechnol. },
  number = {3},
  pages = {e1779},
  publisher = {Wiley Online Library},
  title = {Clinical magnetic hyperthermia requires integrated magnetic particle imaging},
  doi = {10.1002/wnan.1779},
  volume = {14},
  year = {2022}
}

@article{kalwarczyk2011comparative,
  author = {Kalwarczyk, Tomasz and Ziebacz, Natalia and Bielejewska, Anna and Zaboklicka, Ewa and Koynov, Kaloian and Szymanski, Jedrzej and Wilk, Agnieszka and Patkowski, Adam and Gapinski, Jacek and Butt, Hans-Jürgen and others},
  journal = {Nano Lett.},
  number = {5},
  pages = {2157--2163},
  title = {Comparative analysis of viscosity of complex liquids and cytoplasm of mammalian cells at the nanoscale},
  doi = {10.1021/nl2008218},
  volume = {11},
  year = {2011}
}

@book{kim2013microhydrodynamics,
  author = {Kim, Sangtae and Karrila, Seppo J},
  publisher = {Elsevier},
  title = {Microhydrodynamics: principles and selected applications},
  year = {1991},
  doi = {10.1016/B978-0-7506-9173-4.50001-3}
}

@article{kole2020assessing,
  author = {Kole, Koen and Zhang, Yiping and Jansen, Eric JR and Brouns, Terence and Bijlsma, Ate and Calcini, Niccolo and Yan, Xuan and Lantyer, Angelica da Silva and Celikel, Tansu},
  journal = {Nat. Neurosci.},
  number = {9},
  pages = {1044--1046},
  title = {Assessing the utility of Magneto to control neuronal excitability in the somatosensory cortex},
  doi = {10.1038/s41593-019-0474-4},
  volume = {23},
  year = {2020}
}

@article{kwapiszewska2020nanoscale,
  author = {Kwapiszewska, Karina and Szczepa{\'n}ski, Krzysztof and Kalwarczyk, Tomasz and Michalska, Bernadeta and Patalas-Krawczyk, Paulina and Szyma{\'n}ski, Jędrzej and Andryszewski, Tomasz and Iwan, Michalina and Duszy{\'n}ski, Jerzy and Ho{\l}yst, Robert},
  journal = {J. Phys. Chem. Lett.},
  number = {16},
  pages = {6914--6920},
  publisher = {ACS Publications},
  title = {Nanoscale viscosity of cytoplasm is conserved in human cell lines},
  doi = {10.1021/acs.jpclett.0c01748},
  volume = {11},
  year = {2020}
}

@article{latypova2024magnetogenetics,
  author = {Latypova, Anastasiia A and Yaremenko, Alexey V and Pechnikova, Nadezhda A and Minin, Artem S and Zubarev, Ilya V},
  journal = {J. Nanobiotechnology},
  number = {1},
  pages = {327},
  publisher = {Springer},
  title = {Magnetogenetics as a promising tool for controlling cellular signaling pathways},
  doi = {10.1186/s12951-024-02616-z},
  volume = {22},
  year = {2024}
}

@article{leliaert2017adaptively,
  author = {Leliaert, Jonathan and Mulkers, Jeroen and De Clercq, Jonas and Coene, Annelies and Dvornik, M and Van Waeyenberge, Bartel},
  journal = {AIP Adv.},
  number = {12},
  pages = {125010},
  title = {Adaptively time stepping the stochastic Landau-Lifshitz-Gilbert equation at nonzero temperature: Implementation and validation in MuMax3},
  doi = {10.1063/1.5003957},
  volume = {7},
  year = {2017}
}

@Article{mamiya2011hyperthermic,
  author = {Mamiya, Hiroaki and Jeyadevan, Balachandran},
  doi = {10.1038/srep00157},
  fjournal = {Sci. Rep.},
  journal = {Sci. Rep.},
  number = {1},
  pages = {1--7},
  publisher = {Nature Publishing Group},
  title = {Hyperthermic effects of dissipative structures of magnetic nanoparticles in large alternating magnetic fields},
  volume = {1},
  year = {2011}
}

@article{meister2016physical,
  author = {Meister, Markus},
  journal = {eLife},
  pages = {e17210},
  title = {Physical limits to magnetogenetics},
  volume = {5},
  year = {2016},
  doi = {10.7554/eLife.17210}
}

@article{okada2025proposal,
  author = {Okada, Kazuya and Satoh, Akira},
  doi = {10.1016/j.jmmm.2025.173259},
  fjournal = {J. Magn. Magn. Mater.},
  journal = {J. Magn. Magn. Mater.},
  pages = {173259},
  title = {Proposal of hybrid-type simulation techniques for spherical magnetic nanoparticles with uniaxial anisotropy: A combination of Brownian dynamics and Monte Carlo methods achieving fast and scalable simulations},
  volume = {629},
  year = {2025}
}

@article{ota2021empirical,
  author = {Ota, Satoshi and Ohkawara, Seiichi and Hirano, Harutoyo and Futagawa, Masato and Takemura, Yasushi},
  journal = {J. Magn. Magn. Mater.},
  pages = {168354},
  publisher = {Elsevier},
  title = {Empirical and simulated evaluations of easy-axis dynamics of magnetic nanoparticles based on their magnetization response in alternating magnetic field},
  volume = {539},
  year = {2021},
  doi = {10.1016/j.jmmm.2021.168354}
}

@article{ota2017evaluation,
  author = {Ota, Satoshi and Takemura, Yasushi},
  journal = {Appl. Phys. Express},
  number = {8},
  pages = {085001},
  publisher = {IOP Publishing},
  title = {Evaluation of easy-axis dynamics in a magnetic fluid by measurement and analysis of the magnetization curve in an alternating magnetic field},
  volume = {10},
  year = {2017},
  doi = {10.7567/APEX.10.085001}
}

@article{ovejero2021selective,
  author = {Ovejero, Jesus G and Armenia, Ilaria and Serantes, David and Veintemillas-Verdaguer, Sabino and Zeballos, Nicoll and L{\'o}pez-Gallego, Fernando and Gr{\"u}ttner, Cordula and de la Fuente, Jes\'us M and Puerto Morales, Mar{\'\i}a del and Grazu, Valeria},
  doi = {10.1021/acs.nanolett.1c02178},
  journal = {Nano Lett.},
  number = {17},
  pages = {7213--7220},
  publisher = {ACS Publications},
  title = {Selective magnetic nanoheating: Combining iron oxide nanoparticles for multi-hot-spot induction and sequential regulation},
  volume = {21},
  year = {2021}
}

@article{perigo2015fundamentals,
  author = {Perigo, E. A. and Hemery, G. and Sandre, O. and Ortega, D. and Garaio, E. and Plazaola, F. and Teran, F. J.},
  doi = {10.1063/1.4935688},
  journal = {Appl. Phys. Rev.},
  pages = {041302},
  title = {Fundamentals and advances in magnetic hyperthermia},
  volume = {2},
  year = {2015}
}

@article{phong2017study,
  author = {Phong, PT and Nguyen, LH and Phong, LTH and Nam, PH and Manh, DH and Lee, I--J and Phuc, NX},
  journal = {J. Magn. Magn. Mater.},
  pages = {36--42},
  publisher = {Elsevier},
  title = {Study of specific loss power of magnetic fluids with various viscosities},
  volume = {428},
  year = {2017},
  doi = {10.1016/j.jmmm.2016.12.008}
}

@article{reeves2014nonlinear,
  author = {Reeves, Daniel B and Weaver, John B},
  journal = {Appl. Phys. Lett.},
  number = {10},
  pages = {102403},
  title = {Nonlinear simulations to optimize magnetic nanoparticle hyperthermia},
  doi = {10.1063/1.4867987},
  volume = {104},
  year = {2014}
}

@article{ruta2015unified,
  author = {Ruta, Sergiu and Chantrell, R and Hovorka, O},
  doi = {10.1038/srep09090},
  journal = {Sci. Rep.},
  number = {1},
  pages = {9090},
  publisher = {Nature Publishing Group UK London},
  title = {Unified model of hyperthermia via hysteresis heating in systems of interacting magnetic nanoparticles},
  volume = {5},
  year = {2015}
}

@article{sanchez2009rotational,
  author = {S{\'a}nchez, Jorge H and Rinaldi, Carlos},
  journal = {J. Magn. Magn. Mater.},
  number = {19},
  pages = {2985--2991},
  publisher = {Elsevier},
  title = {Rotational Brownian dynamics simulations of non-interacting magnetized ellipsoidal particles in dc and ac magnetic fields},
  doi = {10.1016/j.jmmm.2009.04.066},
  volume = {321},
  year = {2009}
}

@book{satohintroduction,
  author = {Satoh, A},
  publisher = {Elsevier},
  title = {Introduction to Molecular-Microsimulation of Colloidal Dispersions},
  year = {2003}
}

@book{satoh2017book,
  author = {Satoh, Akira},
  publisher = {CRC Press},
  title = {Modeling of magnetic particle suspensions for simulations},
  year = {2017},
  doi = {10.1201/9781315166094}
  
}

@article{saville2014formation,
  author = {Saville, Steven L and Qi, Bin and Baker, Jonathon and Stone, Roland and Camley, Robert E and Livesey, Karen L and Ye, Longfei and Crawford, Thomas M and Mefford, O Thompson},
  doi = {10.1016/j.jcis.2014.03.007},
  journal = {J. Colloid Interface Sci.},
  pages = {141--151},
  title = {The formation of linear aggregates in magnetic hyperthermia: Implications on specific absorption rate and magnetic anisotropy},
  volume = {424},
  year = {2014}
}

@article{serantes2018anisotropic,
  author = {Serantes, David and Chantrell, Roy and Gavil{\'a}n, Helena and del Puerto Morales, Mar{\'\i}a and Chubykalo-Fesenko, Oksana and Baldomir, Daniel and Satoh, Akira},
  doi = {10.1039/C8CP02768D},
  journal = {Phys. Chem. Chem. Phys.},
  number = {48},
  pages = {30445--30454},
  publisher = {Royal Society of Chemistry},
  title = {Anisotropic magnetic nanoparticles for biomedicine: bridging frequency separated AC-field controlled domains of actuation},
  volume = {20},
  year = {2018}
  
}

@article{serantes2014multiplying,
  author = {Serantes, David and Simeonidis, Konstantinos and Angelakeris, Makis and Chubykalo-Fesenko, Oksana and Marciello, Marzia and Morales, Maria Del Puerto and Baldomir, Daniel and Martinez-Boubeta, Carlos},
  doi = {10.1021/jp410717m},
  journal = {J. Phys. Chem. C},
  number = {11},
  pages = {5927--5934},
  publisher = {ACS Publications},
  title = {Multiplying magnetic hyperthermia response by nanoparticle assembling},
  volume = {118},
  year = {2014}
}

@article{simeonidis2016situ,
  author = {Simeonidis, Konstantinos and Morales, M Puerto and Marciello, Marzia and Angelakeris, Makis and de La Presa, Patricia and Lazaro-Carrillo, Ana and Tabero, Andrea and Villanueva, Angeles and Chubykalo-Fesenko, Oksana and Serantes, David},
  journal = {Sci. Rep.},
  number = {1},
  pages = {38382},
  title = {In-situ particles reorientation during magnetic hyperthermia application: Shape matters twice},
  doi = {10.1038/srep38382},
  volume = {6},
  year = {2016}
}

@article{soukup2015situ,
  author = {Soukup, Dalibor and Moise, Sandhya and C{\'e}spedes, Eva and Dobson, Jon and Telling, Neil D},
  journal = {ACS Nano},
  number = {1},
  pages = {231--240},
  publisher = {ACS Publications},
  title = {In situ measurement of magnetization relaxation of internalized nanoparticles in live cells},
  volume = {9},
  year = {2015},
  doi = {10.1021/nn503888j}
}

@Article{suzuki2021behaviour,
  author = {Suzuki, Seiya and Satoh, Akira and Futamura, Muneo},
  doi = {10.1080/00268976.2021.1892225},
  fjournal = {Molecular Physics},
  journal = {Mol. Phys.},
  number = {9},
  pages = {e1892225},
  title = {The behaviour of magnetic spherical particles and the heating effect in a rotating magnetic field via Brownian dynamics simulations},
  volume = {119},
  year = {2021}
}

@article{usov2012dynamics,
  author = {Usov, NA and Liubimov, B Ya},
  journal = {J. Appl. Phys.},
  number = {2},
  pages = {023901},
  publisher = {AIP Publishing},
  title = {Dynamics of magnetic nanoparticle in a viscous liquid: Application to magnetic nanoparticle hyperthermia},
  volume = {112},
  year = {2012},
  doi = {10.1063/1.4737126}
}

@article{utkur2017relaxation,
  author = {Utkur, Mustafa and Muslu, Yavuz and Saritas, Emine Ulku},
  journal = {Phys. Med. Biol.},
  number = {9},
  pages = {3422--3439},
  title = {Relaxation-based viscosity mapping for magnetic particle imaging},
  doi = {10.1088/1361-6560/62/9/3422},
  volume = {62},
  year = {2017}
}

@article{velazquez2025advances,
  author = {Velazquez-Albino, Ambar C and Imhoff, Eric Daniel and Rinaldi-Ramos, Carlos M},
  journal = {Sci. Adv.},
  number = {2},
  pages = {eado7356},
  title = {Advances in engineering nanoparticles for magnetic particle imaging (MPI)},
  doi = {10.1126/sciadv.ado7356},
  volume = {11},
  year = {2025}
}

@article{wang2020revaluation,
  author = {Wang, Guangfu and Zhang, Peng and Mendu, Suresh K and Wang, Yali and Zhang, Yajun and Kang, Xi and Desai, Bimal N and Zhu, J Julius},
  journal = {Nat. Neurosci.},
  number = {9},
  pages = {1047--1050},
  publisher = {Nature Publishing Group US New York},
  title = {Revaluation of magnetic properties of Magneto},
  doi = {10.1038/s41593-019-0473-5},
  volume = {23},
  year = {2020}
}

@Article{wells2021challenges,
  author = {Wells, James and Ortega, Daniel and Steinhoff, Uwe and Dutz, Silvio and Garaio, Eneko and Sandre, Olivier and Natividad, Eva and Cruz, Maria M and Brero, Francesca and Southern, Paul and others},
  doi = {10.1080/02656736.2021.1892837},
  fjournal = {International Journal of Hyperthermia},
  journal = {Int. J. Hyperther.},
  number = {1},
  pages = {447--460},
  title = {Challenges and recommendations for magnetic hyperthermia characterization measurements},
  volume = {38},
  year = {2021}
}

@Article{wolfschwenger2024molecular,
  author = {Wolfschwenger, Manuel and Jaufenthaler, Aaron and Hanser, Friedrich and Gamper, Jakob and Hofer, Thomas S and Baumgarten, Daniel},
  doi = {https://doi.org/10.1016/j.apm.2024.07.031},
  fjournal = {Applied Mathematical Modelling},
  journal = {Appl. Math. Model.},
  pages = {115624},
  publisher = {Elsevier},
  title = {Molecular dynamics modelling of interacting magnetic nanoparticles for investigating equilibrium and dynamic ensemble properties},
  volume = {136},
  year = {2024}
}

@Article{yanes2007effective,
  author = {Yanes, R and Chubykalo-Fesenko, O and Kachkachi, H and Garanin, DA and Evans, R and Chantrell, RW},
  doi = {10.1103/physrevb.76.064416},
  journal = {Phys. Rev. B},
  number = {6},
  pages = {064416},
  title = {Effective anisotropies and energy barriers of magnetic nanoparticles with N{\'e}el surface anisotropy},
  volume = {76},
  year = {2007}
}

@article{Cam2026,
  author = {Çam, Necda and López-Vázquez, Iago and Iglesias, {\`O}scar and Serantes, David},
  journal = {Nanoscale Adv.},
  number = {12},
  pages = {3515-3523},
  title = {Tuning field amplitude to minimise heat-loss variability in magnetic hyperthermia},
  volume = {8},
  year = {2026},
  doi = {10.1039/D6NA00235H}
}
\bibliographystyle{rsc} 

\end{document}